\documentclass[preprint,12pt]{elsarticle}

\usepackage[margin=1in]{geometry}

\usepackage{graphicx}

\usepackage{amssymb}

\usepackage{url}

\usepackage{amsmath}
\usepackage{upgreek}

\usepackage{amsthm}

\def\lsOp#1{\left\llbracket #1 \right\rrbracket_{f\rightarrow p}}
\def\cellToFace#1{\left\langle #1 \right\rangle_{p\rightarrow f_k}}

\usepackage{dsfont}

\usepackage{siunitx}

\usepackage{amssymb}

\usepackage{bbm}

\usepackage{soul}

\usepackage{chemformula}

\usepackage{algorithmicx}
\usepackage{algorithm}
\usepackage[noend]{algpseudocode}
\makeatletter
\renewcommand{\ALG@beginalgorithmic}{\footnotesize}
\makeatother

\usepackage{MnSymbol}
\usepackage{stmaryrd}

\usepackage{booktabs}
\usepackage{makecell}

\usepackage{amsmath}

\usepackage{subfig}
\makeatletter
\newcommand{\vast}{\bBigg@{4}}
\newcommand{\Vast}{\bBigg@{7}}
\makeatother

\newcommand*\xbar[1]{%
   \hbox{%
     \vbox{%
       \hrule height 0.5pt %
       \kern0.5ex%
       \hbox{%
         \kern-0.1em%
         \ensuremath{#1}%
         \kern-0.1em%
       }%
     }%
   }%
}

\usepackage{caption}

\setcitestyle{authoryear}
 \biboptions{comma,round}

\date{}

\journal{Elsevier}

\begin{document}

\begin{frontmatter}

\title{A robust phase-field method for two-phase flows on unstructured grids}
\author{Hanul Hwang\corref{cor1}}
\ead{hanul@stanford.edu}
\author{Suhas S. Jain}
\ead{sjsuresh@stanford.edu}
\cortext[cor1]{Corresponding author}
\address{Center for Turbulence Research, Stanford University, California, USA 94305}

\begin{abstract}
A phase-field method for unstructured grids that is accurate, conservative, and robust is proposed in this work. The proposed method also results in bounded transport of volume fraction, and the interface thickness adapts automatically to local grid size.
In addition to this, we present a novel formulation for two-phase flows on collocated grids that is provably energy stable, which is a critical feature for robust simulations of two-phase turbulent flows.

The proposed method is first evaluated on canonical test cases, including scenarios like drop advection and drop in a shear flow. The accuracy of the proposed method in the presence of grid transitions, which is important for simulations in complex geometries using unstructured grids, is also evaluated. We assess the robustness of our method by performing simulations of a drop in homogeneous isotropic turbulence at infinite Reynolds number with varying density ratios. Furthermore, as a step toward verification and validation, simulations of test cases in complex geometries and conditions, such as, a damped surface wave, infinitesimal interface distortion on a liquid jet, and the atomization of a liquid jet from the engine combustion network's Spray A nozzle are presented. These simulations illustrate the accuracy, robustness, and applicability of the proposed method in various kinds of complex two-phase flow applications of engineering interest.

\end{abstract}


\end{frontmatter}

\section{Introduction\label{sec:intro}}

Two-phase flows are prevalent in nature and industrial applications. Examples include, surface cooling~\citep{liu2019}, wind-wave interaction~\citep{buckley2016}, Rayleigh--Taylor~\citep{sharp1984overview,hwang2021scale,chan2023theory} and Richtmyer--Meshkov~\citep{holmes1999richtmyer} instabilities, atomization of liquid jets~\citep{gorokhovski2008modeling,prakash2019detailed,hwang2022} and droplets~\citep{jain2019secondary}, and flows with phase change~\citep{tryggvason2005direct} and moving contact lines~\citep{sui2014numerical}.
In such flows, computational modeling of interfaces separating two immiscible fluids is numerically challenging due to the discontinuity in quantities such as density and viscosity across the interface.
To tackle this challenge, a variety of interface-capturing and interface-tracking methods have been developed. These include volume-of-fluid (VOF), level-set (LS), phase-field (PF), and front-tracking methods, as well as hybrid methods such as coupled LS and VOF and coupled PF and VOF methods~\citep*{mirj_jain_dodd2017}.
There are strengths and weaknesses to each of the methods. Researchers, however, are actively investigating ways to maintain a sharper and more accurate representation of interface, avoid undershoots and overshoots in volume fraction, and conserve mass (while maintaining computational efficiency).

More recently, the interface-capturing and interface-tracking methods are being extended to unstructured grid settings. The robust, efficient, and accurate implementation of these methods on unstructured grids is crucial for the simulations of two-phase flows in complex geometries for practical engineering applications.
The use of unstructured grids can result in relatively easier and automatic grid generation without manual intervention in complex geometries; and it can also reduce the total number of cells in the computational domain due to the flexibility in using coarser grids away from the regions of interest. Due to the easy of domain decomposition, the mesh generation is also fast and scalable.
The VOF method has been widely used for simulations of two-phase flows, since it is known to conserve mass and have advantages in handling complex interface morphology changes. The VOF method can be classified into geometric and algebraic types. Geometric VOF reconstructs the interface based on geometric representation (e.g., plane or quadratic surface), and the reconstructed segments are advected in a Lagrangian fashion~\citep{youngs1982,rider1998}. 
Accordingly, geometric VOF method maintains sharp interfaces; however, its geometric manipulations become inherently complicated on unstructured grids. Moreover, operations related to interfaces are localized in the interface cells, leading to high computational cost and parallel load imbalance~\citep{jofre2015}. Nevertheless, this method has been successfully extended to unstructured grids~\citep{ivey2015accurate,ivey2017conservative,xie2017}.

Various implementations of the geometric VOF on unstructured grids have been proposed in the literature. \citet{Mosso1997} and \citet{Kothe1999} introduced a second-order method for unstructured grids using remapping technique. \citet{Shahbazi2003} extended the piecewise linear interface calculation (PLIC)-based method for triangular meshes and proposed the geometric least squares method for computing normals on unstructured grids. \citet{Ashgriz2004} extended the least squares fit method of~\citet{Scardovelli2003} for normal calculation in a PLIC-type method for unstructured grids. \citet{Dyadechko2005} applied the moment-of-fluid (MOF) approach, naturally suited for unstructured grids without special treatment. \citet{yang2006numerical} developed an analytic formulation for interface reconstruction on triangular and tetrahedral meshes, resulting in faster interface reconstruction compared to iterative methods. \citet{Ahn2007} introduced multi-material VOF and MOF methods for unstructured grids. \citet{Mosso2009} proposed the patterened interface reconstruction algorithm that reduces the discontinuities between interfaces in neighboring cells. \citet{Ito2013} extended the PLIC method for arbitrary unstructured grids and the reconstructed distancing function algorithm~\citep{cummins2005estimating} for improved calculation of curvature on unstructured grids. \citet{Maric2013} devised an unsplit geometric VOF method for unstructured grids with local dynamic adaptive mesh refinement. \citet{Ito2014} developed an unstructured grid height function method for calculating normals and curvature. \citet{Jofre2014} suggested a three-dimensional (3D) extension of the unsplit method for unstructured grids. \citet{evrard2017estimation} developed a method for curvature calculation using parabolic reconstruction of the interface on unstructured grids. \citet{Patel2017} formulated a balanced force algorithm using hybrid staggered/non-staggered formulation on unstructured grids.

On the other hand, the algebraic VOF method computes the face fluxes algebraically using compressive difference schemes, avoiding reconstruction of the interface and related issues~\citep{lafaurie1994}. The absence of geometric reconstruction makes these methods attractive over geometric VOF methods for unstructured grids.
\citet{Ubbink1999} developed the compressive interface capturing scheme for arbitrary meshes (CISCSAM) method, and showed its applicability on unstructured grids.
A detailed evaluation of another algebraic VOF method in the open-source solver \texttt{interFOAM} can be found in \citet{Deshpande2012}.
Nonetheless, an issue related to the smearing of the interface persisted, especially, in the region of high shear and high local Courant–Friedrichs–Lewy (CFL) number. 
To address this smearing issue of the interface, \cite{Darwish2006} later developed the switching
technique for advection and capturing of surfaces (STACS) scheme, which preserves the interface sharpness for all CFL numbers.
More recently, \citet{Roenby2016} developed a method based on the concept of isosurfaces for unstructured grids and showed improvement over compressive algebraic VOF schemes.

Furthermore, tangent of hyperbolic interface capturing (THINC) method, a hybrid geometric/algebraic approach, was proposed ~\citep{xiao2005,xie2017} to overcome the numerical issues of VOF methods. However, its application to unstructured grid adds another level of complexity ~\citep{Ii2014,xie2017}. Additionally, it is reported that the phase indicator variable is not bounded~\citep{kim2021,kumar2021thinc} without the use of flux limiters.
As one of the alternative choices of interface-capturing schemes, it is worth mentioning the level-set formulations on unstructured grids. \cite{Kees2011} and \cite{Balcazar2014} developed a conservative level-set method (CLS) on unstructured grids, and \cite{Antepara2021} extended the CLS method for adaptive unstructured grids. More recently, \cite{Janodet2022} extended the accurate CLS (ACLS) method for adaptive unstructured grids. However, the level-set methods are becoming less popular for the simulation of two-phase flows due to the inherent issue of non conservation of mass/volume in these methods.
Interface-tracking methods, such as a front-tracking method \citep[see][]{Hua2019}, have also been developed on unstructured grids. However, the front-tracking methods are less popular due to the need for special treatment for topological changes such as coalescence and breakup.
Various hybrid methods have also been formulated on unstructured grids. \citet{Lv2010} developed a coupled LS and algebraic VOF method based on bounded compressive scheme for 3D tetrahedral grids; \cite{Balcazar2016,Cao2017,Ferrari2017} developed coupled VOF and LS methods; and \citet{maric2015lentfoam} and \cite{Liu2021} developed hybrid LS and front-tracking methods, to name a few.
Recently, the phase-field (PF) methods have been gaining popularity as an alternative interface-capturing method with many attractive features. The PF method or a diffuse-interface (DI) method models the interface as a diffused layer over several cells, as the name indicates, rather than in a sharp discontinuous representation. Thus, numerical operations, such as derivatives, can be applied even across the interface, as is done in a single-phase problem, leading to high scalability of the method.
Moreover, the sharpness of the interface is easily controlled by the use of an interface-regularization term. 
There are various PF methods in the literature, which can be classified into Cahn-Hilliard-based methods \citep{cahn1958free} and Allen-Cahn-based methods \citep{allen1979microscopic}. A Cahn-Hilliard PF model is conservative but involves a fourth-order spatial derivative in the equation, which requires careful construction of the numerical methods. In contrast, an Allen-Cahn PF model does not involve fourth-order derivatives in the equation and is not conservative. A conservative version of the Allen-Cahn based model has been proposed by \citet{chiu2011conservative}. This model has been extended to unstructured grids in the context of a finite-element formulation by \cite{joshi2018adaptive}. A flux-limiter-based diffuse-interface approach for compressible flows on unstructured grids has been developed by \cite{chiapolino2017sharpening}.
More recently, \citet{jain2022accurate} proposed an accurate conservative diffuse-interface/phase-field (ACDI/ACPF) model and showed that this method is robust, conservative, and significantly more accurate than other PF models in the literature. \citet{jain2022accurate} showed that this PF model can maintain boundedness of the volume fraction while maintaining the interface thickness on the order of only one or two grid points using low-dissipative central schemes. The capability of this PF model to maintain such sharp interfaces without the need for any special geometric treatment, unlike the VOF method, makes it a highly attractive interface-capturing method for accurate simulations of two-phase flows at an affordable cost.

The ACDI method introduces an algebraic signed-distance-like auxiliary function replacing the PF variable in the sharpening term of the interface-regularization term. Thus, evaluation of (sharp) gradients in the PF variable across interfaces is eliminated, yielding a more accurate computation of the sharpening flux. As a result, the ACDI method demonstrates enhanced accuracy compared to the previously reported conservative PF methods and imposes a less restrictive CFL condition. Moreover, it improves the computation of the surface tension force by using this auxiliary function to compute the curvature and significantly reduces the spurious velocity fields near interfaces.
Altogether, this PF method has great potential as an accurate, cost-efficient, and highly scalable interface-capturing method for the simulation of complex interfacial flows. Therefore, this PF model will be used in this work and extension of this model for unstructured grids will be proposed. We have also extended the conservative PF model of \citet{chiu2011conservative} to unstructured grids, and the comparison of CDI and ACDI models on an unstructured grid is presented in~\ref{sec:app:cdi comparison}.
For accurate numerical simulations of turbulent flows at high Reynolds numbers, it is known that one needs to use low-dissipative numerical schemes that conserve global kinetic energy (in the absence of viscous dissipation and time-stepping errors), particularly in the context of large-eddy simulations (LES). Along these lines \citet{perot2000conservation,zhang2002accuracy} developed a staggered grid formulation on unstructured grids that achieve discrete energy conservation. More recently, \citet{Jofre2014} derived a staggered grid formulation for two-phase flows on unstructured grids and showed that the kinetic energy is discretely conserved. However, implementation of a staggered grid formulation can be quite challenging, particularly on unstructured grids. Hence, a collocated grid formulation is a preferred choice because of its ease of implementation. 

\citet{morinishi1998fully} showed that the pressure term in the collocated grid formulation introduces an error in kinetic energy transport for single-phase flows. The error was shown to be strictly dissipative in nature, and hence the formulation is energy stable. \citet{felten2006kinetic} found that this dissipative error was not large and did not adversely affect the solution in LES of turbulent flows. However, in the context of two-phase flows, \citet{Jofre2014} found that the kinetic energy could increase with time in the collocated grid formulation for two-phase flows, which could make the formulation non-robust. To the best of our knowledge, there is no provably energy stable formulation on collocated grids for two-phase flows. In this work, we propose a novel consistent formulation on collocated grids for two-phase flows and show that the error associated with the pressure terms in collocated grid formulation is strictly dissipative in nature even for two-phase flows. Hence, the proposed formulation on collocated grid is provably energy stable, and therefore, robust for simulations of two-phase flows.  
The objective of the current work is to propose and extend the recently developed ACDI method for unstructured grids that will allow for accurate simulations of two-phase flows in complex geometries and for large-scale practical applications in engineering. We propose an implementation of this phase-field model on unstructured Voronoi grids and evaluate its performance using various interface advection test cases and hydrodynamic-coupled simulations.
We also propose a robust formulation for simulation of two-phase flows on collocated grids. 
On collocated grids, it is known that the global kinetic energy is not conserved in the absence of dissipative mechanisms due to the errors introduced by the pressure terms. 
With the correct choice of construction of the mass flux at the interface, we can show that the error associated with the pressure terms in a fractional step formulation is strictly dissipative in nature for two-phase flows. To the best of our knowledge, the proposed formulation in this work is the first provably energy-stable formulation for two-phase flows on collocated grids.
We also design and propose new standardized test cases for the evaluation of interface-capturing methods on unstructured grids.
These test cases help systematically evaluate the errors due to grid transitions that are commonly encountered on unstructured grids. At the end, we also present more complex simulations, including an atomization and spray formation of a high-density ratio liquid jet from a realistic injector.

The rest of this paper is organized as follows.
Section~\ref{sec:two-phase formulation} introduces the accurate conservative diffuse-interface model and the consistent governing equations.
Section~\ref{sec:voronoi grids and discretization} presents a Voronoi mesh-based unstructured grid framework and discretization of the phase-field model.
Section~\ref{sec:projection method algorithm} provides a detailed description of the projection method algorithm. The derivation of the discrete energy conservation equations in both staggered and collocated framework is presented in Section~\ref{sec:discrete energy conservation}.
Section~\ref{sec:results} and Section~\ref{sec:complex two phase flow apps} present verification and validation test cases, focusing on canonical and complex cases, respectively.
Finally, concluding remarks are presented in Section~\ref{sec:conclusions}.

\section{Two-phase formulation\label{sec:two-phase formulation}}
\subsection{Accurate conservative phase-field/diffuse-interface model}
In this work, we consider incompressible two-fluid flows. The volume fraction of one of the fluids is indicated by the phase-field variable $\phi=\phi_1$, which satisfies the relation $\sum_{l=1,2}\phi_l$=1 by definition, where the subscript $l$ denotes phase index. Then, the ACDI model for $\phi$ \citep{jain2022accurate} can be written as
\begin{equation}
\frac{\partial \phi}{\partial t} + \vec{\nabla}\cdot(\phi \vec{u}) = \vec{\nabla}\cdot\left\{\Gamma\left\{\epsilon\vec{\nabla}\phi - \frac{1}{4} \left[1 - \tanh^2{\left(\frac{\psi}{2\epsilon}\right)}\right]\frac{\vec{\nabla} \psi}{|\vec{\nabla} \psi|}\right\}\right\},
\label{eq:acdi}
\end{equation}
where $\vec{u}$ is the velocity, $\Gamma$ denotes the velocity-scale parameter, and $\epsilon$ is the interface thickness scale parameter. 
Here, $\psi$ is an auxiliary variable and represents the signed-distance-like function from the interface, which is defined as
\begin{equation}
    \psi = \epsilon \ln\left(\frac{\phi + \varepsilon}{1 - \phi + \varepsilon}\right).
    \label{eq:psi}
\end{equation}
Note that a small number $\varepsilon$ is added to both the numerator and denominator to 
avoid division by 0, and $\varepsilon$ is chosen as $\varepsilon = 10^{-100}$.
The right-hand-side (RHS) term of Eq.~\eqref{eq:acdi} is the interface-regularization term, which is an artificial term that is added to maintain the constant thickness of the interface. The interface-regularization term contains a diffusion and a sharpening term, which balance each other to maintain a constant interface thickness on the order of $\epsilon$.

\citet{jain2022accurate} proposed that $\phi$ remains bounded between $0$ and $1$ using a central scheme, provided $\Gamma/|\vec{u}|_{max}\ge 1$ and $\epsilon/\Delta_f > 0.5$ conditions are met, where $\Delta_f$ represents the local grid size. In Section \ref{sec:results}, we verify that the same condition is true on unstructured grids for maintaining boundedness of $\phi$.

\subsection{Consistent momentum transport \label{sec:consistent mom}}
The addition of the regularization term in Eq.~\eqref{eq:acdi} modifies the way the volume fraction is transported. This modified transport of volume fraction should be consistently accounted for in the transport of mass, momentum, and energy \citep{jain2022accurate}. 
If we represent the artificial regularization volume flux as $\vec{a}$,
\begin{equation}
    \vec{a} = \Gamma\left\{\epsilon\vec{\nabla}\phi - \frac{1}{4} \left[1 - \tanh^2{\left(\frac{\psi}{2\epsilon}\right)}\right]\frac{\vec{\nabla} \psi}{|\vec{\nabla} \psi|}\right\},
    \label{eq:volume flux}
\end{equation}
then the implied artificial mass flux $\vec{f}$ can be written as
\begin{equation}
    \vec{f}=\sum_{l=1}^2 \rho_l \vec{a}_l = \left\{ (\rho_1-\rho_2) \Gamma\left\{\epsilon\vec{\nabla}\phi - \frac{1}{4} \left[1 - \tanh^2{\left(\frac{\psi}{2\epsilon}\right)}\right]\frac{\vec{\nabla} \psi}{|\vec{\nabla} \psi|}\right\} \right\},
    \label{eq:artificial mass flux}
\end{equation}
where $\rho_l$ is the density of phase $l$.
As a result, the mass transport equation has an additional term due to the artificial mass flux, which is written as
\begin{equation}
\frac{\partial \rho}{\partial t} + \frac{\partial \rho u_j}{\partial x_j} = \frac{\partial f_j}{\partial x_j},
\label{eq:mod_continuity}
\end{equation}
where the density $\rho$ can be expressed in terms of volume fractions as $\rho = \rho_1\phi + \rho_2(1-\phi)$.

Furthermore, the effect of the regularization term is considered in the momentum equation as
\begin{equation}
\frac{\partial \rho u_i}{\partial t} + \frac{\partial \rho u_i u_j}{\partial x_j} =  -\frac{\partial p}{\partial x_i} + \frac{\partial \tau_{ij}}{\partial x_j} + \frac{\partial u_i f_j}{\partial x_j} + \sigma \kappa \frac{\partial \phi_1}{\partial x_i},  %
\label{eq:momf}
\end{equation}
where $i$ and $j$ are the Einstein indices, $\tau_{ij}$ is the stress tensor, $\tau_{ij} = 2\mu S_{ij} = \mu(\partial u_i / \partial x_j+\partial u_j / \partial x_i)$, and $p$ is the pressure.

\subsection{Surface tension modeling}

The surface tension forces (last term in Eq. \eqref{eq:momf}) are modeled using a continuum-surface force (CSF) formulation ~\citep{brackbill1992continuum}, where
$\sigma$ is the surface tension coefficient, and $\kappa$ is the surface curvature, defined as $\kappa = -\vec{\nabla}\cdot\vec{n}=-\vec{\nabla}\cdot(\vec{\nabla}\psi/|\vec{\nabla}\psi|)$. The use of $\psi$ to compute curvature in the surface tension force was shown in \cite{jain2022accurate} to improve modeling of the surface tension force and reduce spurious currents significantly. This \textit{improved CSF} model will result in lower spurious currents compared to other existing models in the literature, including the free-energy based surface tension models, and can be easily adopted for unstructured grids. Hence, the \textit{improved CSF} formulation is used to model surface tension forces in this work.

\section{Voronoi grid and discretization\label{sec:voronoi grids and discretization}}
In this work, a Voronoi mesh-based unstructured grid is adopted. A Voronoi tessellation divides the volume of space into regions, where each location in the region is closer to the associated seed point than to any other seed points, based on Euclidean distance.
The grid is generated using the \texttt{stitch} tool, a fast Voronoi meshing package that was developed at Cascade Technologies~\citep{bres2018}, which is now part of Cadence Design Systems. This package generates a hexagonal close-packed (HCP) array of points, and the resulting mesh has uniform 14-sided polyhedral cells in the regions of the domain away from the boundaries and grid transitions.
Figure \ref{fig:unstructured grid} shows a schematic of an unstructured Voronoi grid. 

Some of the advantages of the Voronoi grids, compared to other unstructured grids, are as follows:
\begin{enumerate}
    \item In Voronoi grids, vertex locations uniquely define the geometric information (cell volume, face area, and normals) and the connectivity of the Voronoi mesh (neighbors of a cell). This deterministic connection between the points simplifies mesh generation and manipulation. The quality of mesh can also be improved using Lloyd iteration in a robust and stable way \citep{ambo2020aerodynamic}.
    \item The face normals and cell displacement vectors are parallel to one another by construction. This orthogonality enables computational efficiencies for both the mesh generator and the fluid solver (see Figure~\ref{fig:unstructured grid}).
    \item The cell faces are located at the center of the cell displacement vectors (see Figure~\ref{fig:unstructured grid}), which will make the Voronoi grid suitable for implementation of low-dissipative central schemes.
    \item Voronoi mesh generation is highly parallelizable due to its local nature, and it avoids the issue of degenerate point sets with the Delaunay triangulation \citep{bres2018}.%
\end{enumerate}

\subsection{Discretization of the phase-field model \label{sec:ACDI-discrete}}

In this work, we use a collocated grid arrangement; in other words, all the variables being solved ($\phi$, $\psi$, $\rho$, $u_i$, and $p$) are stored at the Voronoi seed points. This is done to avoid the need to develop complicated operators \citep{perot2000conservation} in an unstructured, staggered grid setting.
The governing equations, Eq.~\eqref{eq:acdi} and Eq.~\eqref{eq:momf}, are spatially discretized using a skew-symmetric-type second-order flux-split central difference scheme \citep{jain2022kinetic}, and a fourth-order Runge-Kutta (RK4) method is chosen as the time-integration scheme.

\begin{figure}
    \centering
    \includegraphics[width=0.5\textwidth]{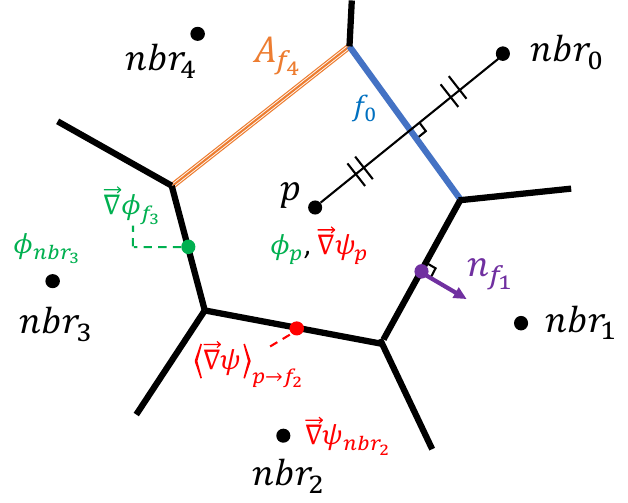}
    \caption{A schematic of an unstructured Voronoi grid. The black dots denote the seed points of cell $p$ and those of its neighboring cells ($nbr_k$). The cell face and the cell displacement vector between the cell $p$ and its neighbor $nbr_k$ are indicated as $f_k$ and $\mathbf{x_p}-\mathbf{x_{nbr_{k}}}$, respectively.}
    \label{fig:unstructured grid}
\end{figure}
The RHS of Eq.~\eqref{eq:acdi} is a divergence of the artificial regularization volume flux $\vec{a}$ in Eq.~\eqref{eq:volume flux}; thus, it is computed in a finite-volume formulation as
\begin{equation}
    \sum_{f_k} \left( \vec{a}_{f_{k}} \cdot \vec{n}_{f_{k}}\right)A_{f_{k}},
    \label{eq:discretization}
\end{equation}
where $f_k$ indicates the cell face between the cell $p$ and that of the neighboring cell $nbr_k$. The $\vec{n}_{f_{k}}$ and $A_{f_{k}}$ are the unit face normal and the cell face area of $f_k$, respectively, as shown in Figure~\ref{fig:unstructured grid}. Because the dot product in Eq.~\eqref{eq:discretization} is calculated at the cell faces, the volume flux $\vec{a}$ must be evaluated at the cell faces. 

The discretization of the diffusion and sharpening terms in Eq.~\eqref{eq:volume flux} in practice is the following. First, the diffusion flux is evaluated at the cell face. Since $\phi$ is stored at the cell center (seed point), the gradient of $\phi$ dotted with the face normal is computed, following Eq.~\eqref{eq:discretization}, as 
\begin{equation}
    \sum_{f_k} \left( \epsilon \vec{\nabla}\phi_{f_{k}} \cdot \vec{n}_{f_{k}}\right)A_{f_{k}} = \sum_{f_k} \epsilon \left\{ (\phi_{nbr_{k}}-\phi_{p})/|\mathbf{x_p}-\mathbf{x_{nbr_{k}}} | \right\} A_{f_{k}},
    \label{eq:diffusion term}
\end{equation}
where $\mathbf{x_p}$ and $\mathbf{x_{nbr_k}}$ denote the displacement vectors of cells $p$ and $nbr_k$, respectively.

Note that, we use the dimensionless parameter $\epsilon^*$ defined as $\epsilon^* = \epsilon / \Delta_{f}$, instead of $\epsilon$, and we prescribe $\epsilon^*$ as a solver input.
Here, $\Delta_{f}$ is a characteristic length of the local grid, e.g., simply the grid size for a uniform mesh. For an unstructured grid with non-uniform mesh size, we propose to use $\Delta_{f} = \Delta_{\text{local}} = |\mathbf{x_p}-\mathbf{x_{nbr_{k}}}|$ to evaluate Eq.~\eqref{eq:diffusion term} at cell faces. As a result, the evaluation of Eq.~\eqref{eq:diffusion term} in practice is
\begin{equation}
    \sum_{f_k} \left( \epsilon \vec{\nabla}\phi_{f_{k}} \cdot \vec{n}_{f_{k}}\right)A_{f_{k}} = \sum_{f_k} \epsilon^* (\phi_{nbr_{k}}-\phi_{p}) A_{f_{k}},
    \label{eq:diffusion term discretization}
\end{equation}
where the final expression no longer includes the local grid size $\Delta_f$.
Next, we compute the sharpening term, which is the second term on the RHS of Eq.~\eqref{eq:acdi}.
This term involves evaluation of $\psi$ and $\vec{\nabla}\psi$ at the cell faces due to the divergence form of the regularization volume flux $\vec{a}$ in Eq.~\eqref{eq:volume flux}.
Note that $\epsilon$ can be removed from the numerator and denominator of the argument of hyperbolic tangent function as well as in computation of interface normal. To compute the sharpening term, in practice, we define a modified auxiliary variable $\widetilde{\psi}$, that is independent of $\epsilon$, as 
\begin{equation}
    \widetilde{\psi} = \frac{\psi}{\epsilon} = \ln\left(\frac{\phi + \varepsilon}{1 - \phi + \varepsilon}\right).
    \label{eq:auxiliary variable psi tilde}
\end{equation}
Then, the sharpening term is calculated as
\begin{equation}
    \sum_{f_k} \frac{1}{4}\left\{ 1-\tanh^2 \left(\frac{\langle \widetilde{\psi}_{p} \rangle_{p\rightarrow f_{k}} }{2} \right) \right\} \left(  \cellToFace{\frac{\vec{\nabla}\widetilde{\psi}_{p}}{|\vec{\nabla}\widetilde{\psi}_{p}|}} \cdot \vec{n}_{f_{k}} \right) A_{f_{k}},
    \label{eq:sharpening term with auxilary variable}
\end{equation}
where the operator $\cellToFace{\cdot}$ is a cell-to-face interpolation, which results in a midpoint value.

The introduction of this additional auxiliary variable $\widetilde{\psi}$ removes the variable $\epsilon$ in the sharpening term. For non-uniform grids, where grid size can vary spatially, Eqs.~\eqref{eq:diffusion term discretization}--\eqref{eq:sharpening term with auxilary variable} simplify the evaluation of the regularization term since it no longer depends on the local grid spacing.
This discretization results in maintaining boundedness of volume fraction without the need for any special treatment.

\section{Projection method algorithm\label{sec:projection method algorithm}}
In this section, the projection method algorithm used in this work is presented in full detail.
Although all the results shown in this work are computed using the RK4 time-stepping scheme, a single full time step from $n$ to $n+1$ is explained using the explicit Euler framework for the sake of simplicity and clarity. In addition, we omit the viscous term here for the same reason.

\paragraph{PF variable and density update}
First, we update $\phi$ by solving
\begin{equation}
   \frac{\phi^{n+1}_{p} - \phi^{n}_{p}}{\Delta t}V_{cv} = - \left[ \sum_{f_{k}} \cellToFace{\phi_p} U_{f_{k}} - \sum_{f_{k}} (a_{j,{f_k}} n_{j,{f_k}}) \right]^n A_{f_{k}},
   \label{eq:algorithm-acdi}
\end{equation}
where the superscripts $n$ and $n+1$ represent time-step indices, 
$V_{cv}$ and $A_{f_{k}}$ are cell volume and face area, respectively,
and $U_{f_k}$ is the mass flux computed at cell face $f_k$ using Eq.~\eqref{eq:algorithm-face flux} below at the previous time step/sub step, which satisfies the divergence-free condition. 
Then, the density in the next time step is updated using $\rho^{n+1} = \rho_1 \phi^{n+1} + \rho_2 (1-\phi^{n+1})$.

\paragraph{Momentum equation update}
We advance the momentum equation to compute the intermediate momentum $\hat{\Phi}_{i,p} = \widehat{\rho u_i}$ at the cell center using the previous time-step value, as
\begin{align}
   \frac{\hat{\Phi}_{i,p} - (\rho^n u^n_i)_{p}}{\Delta t} V_{cv} =
   \alpha \rho^{n+1} \lsOp{
   \left[  -\frac{1}{\cellToFace{\rho_p}} \left(\frac{\partial p}{\partial n}\right)_{f_k} + \frac{1}{\cellToFace{\rho_p}} \left( F_{j,f_{k}} n_{j,f_{k}}\right)  \right]^n
   } \nonumber \\
   - \left[ \sum_{f_{k}} 
   \cellToFace{\rho_p} \cellToFace{u_{i,p}} U_{f_{k}}
   - \sum_{f_{k}} \cellToFace{u_{i,p}} (f_{j,{f_k}} n_{j,{f_k}}) \right]^n A_{f_{k}},
   \label{eq:algorithm-momentum prediction}
\end{align}
where $\alpha$ is a constant that satisfies $0 \leq \alpha \leq 1$, which takes a portion of pressure gradient and external body force terms to compute the intermediate momentum~\citep{francois2006balanced}. The double-lined square bracket in Eq. \eqref{eq:algorithm-momentum prediction} denotes a face-to-cell interpolation using a least square method (details of this least square interpolation are provided in Section~\ref{sec:gradient reconstruction scheme}).

For the convection term of the momentum equation, note that a skew-symmetric-type second-order flux-split central difference scheme is employed as in~\citet{jain2022accurate}. As a result, face interpolated density and velocity are multiplied at cell faces. In fact, it can be shown that this flux-split form of the discretization conserves discrete global kinetic energy (see, Eq.~\eqref{eq:energy budget derivation: discrete form 4} in Section~\ref{sec:energy-stable collocated grid formulation}).

\paragraph{Intermediate face velocity/flux update}
Once the intermediate momentum is updated using Eq.~\eqref{eq:algorithm-momentum prediction}, the so-called predictor step, $\hat{\Phi}_{i,p}$ is corrected to satisfy the divergence-free velocity field condition (also known as the corrector step). This remaining step can be written, in a continuous form, as
\begin{equation}
    \frac{\rho^{n+1}_{p}u^{n+1}_{i,p}-\hat{\Phi}_{i,p}}{\Delta t} =
    \left[-\left(\frac{\partial p}{\partial x_i}\right)_{p} + F_{i,p}\right]^{n+1}.
    \label{eq:algorithm-continous framework-intermediate momentum update 1}
\end{equation}
Here in Eqs. \eqref{eq:algorithm-continous framework-intermediate momentum update 1}-\eqref{eq:algorithm-discrete divergence of the updated velocity}, we temporarily omit (for simplicity of illustration) the contribution of the pressure gradient and external forces from the previous time step, which is the first term on the RHS of Eq.~\eqref{eq:algorithm-momentum prediction}.
Then dividing Eq.~\eqref{eq:algorithm-continous framework-intermediate momentum update 1} by the updated density $\rho^{n+1}_{p}$ and rearranging it for the cell-centered velocity $u^{n+1}_{i,p}$ yields
\begin{equation}
    u^{n+1}_{i,p} =
    \frac{\hat{\Phi}_{i,p}}{\rho^{n+1}_{p}}
    + \frac{\Delta t}{\rho^{n+1}_{p}}
    \left[-\left(\frac{\partial p}{\partial x_i}\right)_{p} + F_{i,p}\right]^{n+1}.
    \label{eq:algorithm-continous framework-intermediate momentum update 2}
\end{equation}
Then, we take discrete divergence and volume integrate Eq.~\eqref{eq:algorithm-continous framework-intermediate momentum update 2}, which will yield
\begin{equation}
    \sum_{f_k} U^{n+1}_{f_k} A_{f_k}= \sum_{f_k} \left\{ \cellToFace{
    \frac{\hat{\Phi}_{i,p}}{\rho^{n+1}_{p}}
    + \frac{\Delta t}{\rho^{n+1}_{p}}
    \left[-\left(\frac{\partial p}{\partial x_i}\right)_{p} + F_{i,p}\right]^{n+1}
    } A_{f_k} \right\} \vec{n}_{i,f_k},
    \label{eq:algorithm-discrete divergence of the updated velocity}
\end{equation}
where $U^{n+1}_{f_k}$ is the updated face velocity. 

It is well known that a collocated formulation leads to velocity/pressure decoupling, which results in  a pressure checkerboard issue. To avoid this issue in the pressure field, we perform Rhie-Chow-like interpolation with a balanced-force approach when updating the RHS of Eq.~\eqref{eq:algorithm-discrete divergence of the updated velocity} as
\begin{equation}
    \sum_{f_k} U^{n+1}_{f_{k}} A_{f_k} = \sum_{f_k}\hat{U}_{f_{k}} A_{f_k} + \sum_{f_k} \frac{\Delta t}{\cellToFace{\rho^{n+1}_{p}}} \left[ -\left(\frac{\partial \delta p}{\partial n}\right)^{n+1}_{f_k} \right] A_{f_k},
    \label{eq:algorithm-update face flux reduced form}
\end{equation}
where we define $\hat{U}_{f_{k}}$ as
\begin{align}
\hat{U}_{f_{k}} = 
\frac{\cellToFace{\hat{\Phi}_{i,p}}}{\cellToFace{\rho^{n+1}_p}} n_{i,f_k}
-\alpha \cellToFace{ \lsOp{
\left[-\frac{1}{\cellToFace{\rho_p}} \left(\frac{\partial p}{\partial n}\right)_{f_k} + \frac{1}{\cellToFace{\rho_p}}\left(F_{j,f_{k}} n_{j,f_{k}}\right) \right]^n
}
} n_{i,f_k}  \nonumber \\
+ \frac{\Delta t}{\cellToFace{\rho^{n+1}_{p}}} \left[ -\left(\frac{\partial p}{\partial n}\right)^n_{f_k} + \left(F_{j,f_k} n_{j,f_{k}}\right)^{n+1}\right].
\label{eq:algorithm-update face flux}
\end{align}
All the terms at cell faces are evaluated consistently, where the numerator is computed as an average from two neighboring cell centers which is divided by the face interpolated density, that is typically what is done in the literature~\citep{Renardy2002,francois2006balanced}. Consequently, the first term on the RHS of Eq.~\eqref{eq:algorithm-update face flux} is a density-weighted velocity rather than just an average of intermediate cell-centered velocity. In addition to the consistent cell-to-face interpolation with the rest of the terms in Eq.~\eqref{eq:algorithm-update face flux}, this choice of density-weighted velocity, $\cellToFace{\hat{\Phi}_{i,p}}/\cellToFace{\rho^{n+1}_p}$, at cell faces to evaluate the discrete divergence of cell-centered velocity $u_{i,p}$ is further justified below in Section~\ref{sec:justification of face velocity}. 

\paragraph{Pressure Poisson solve}
After solving Eqs.~\eqref{eq:algorithm-momentum prediction} and \eqref{eq:algorithm-update face flux}, the intermediate mass flux at the cell faces $\hat{U}_{f}$  contains the contributions from convection terms, the pressure gradient at time step $n$, and other body forces, such as surface tension and gravity force. At this point, the intermediate velocity $\hat{U}_{f_{k}}$ does not necessarily satisfy the divergence-free condition. To make this intermediate velocity divergence free (steps described below in Eqs.~\eqref{eq:algorithm-cell center velocity}--\eqref{eq:algorithm-face flux}) we first solve the pressure Poisson equation to compute the correction pressure $\delta p$ (see Eq.~\eqref{eq:algorithm-pressure correction} for the definition of $\delta p$). The pressure Poisson equation is given by
\begin{equation}
\sum_{f_{k}} \hat{U}_{f_k} A_{f_{k}} = \Delta t \sum_{f_{k}} \frac{1}{\cellToFace{\rho^{n+1}_{p}}} \left(\frac{\partial \delta p}{\partial n}\right)^{n+1}_{f_k} A_{f_{k}}.
\label{eq:algorithm-pressure Poisson}
\end{equation}
Solving Eq.~\eqref{eq:algorithm-pressure Poisson}, we obtain the correction pressure at the cell center $\delta p^{n+1}_p$.

\paragraph{Pressure update}
Next, the pressure at time step $n$ is updated using the correction pressure $\delta p^{n+1}_p$ to obtain the cell center pressure at time step $n+1$, as
\begin{equation}
    p^{n+1}_{p} = p^n_{p} + \delta p^{n+1}_{p}.
    \label{eq:algorithm-pressure correction}
\end{equation}

\paragraph{Cell velocity update}
The pressure gradient and body force at time step $n+1$ are added to the cell velocity to obtain an approximately divergence-free velocity field at the cell center at time step $n+1$, as
\begin{align}
u_{i,p}^{n+1} = \frac{\hat{\Phi}_{i,p}}{\rho^{n+1}_p}
- \alpha \lsOp{
\left[  -\frac{1}{\rho^{n+1}_p} \left(\frac{\partial p}{\partial n}\right)_{f_k} + \frac{1}{\cellToFace{\rho_p}} \left( F_{j,f_{k}} n_{j,f_{k}}\right)  \right]^n
}  \nonumber \\
+ \Delta t \lsOp{
\left[-\frac{1}{\cellToFace{\rho_p}}\left(\frac{\partial p}{\partial n}\right)_{f_k} + \frac{1}{\cellToFace{\rho_p}} \left( F_{j,f_{k}} n_{j,f_{k}}\right) \right]^{n+1}
}.
\label{eq:algorithm-cell center velocity}
\end{align}
Note that $u_{i,p}$ is approximately divergence free since the pressure Poisson equation is solved at cell faces with the intermediate face velocity $\hat{U}_{f_{k}}$. We emphasize that the pressure gradient and the additional forces here are computed at the cell center using the least-square gradient operator (see, Section~\ref{sec:gradient reconstruction scheme}) that uses the values at cell faces. This is done to enforce a discrete balance between the pressure gradient and other forces even at the cell center.

\paragraph{Face velocity/flux update}
Finally, the face flux is updated such that the velocity at the face exactly satisfies the divergence-free condition. Face velocity can be corrected using
\begin{equation}
    U^{n+1}_{f_{k}} = \hat{U}_{f_{k}} + \frac{\Delta t}{\cellToFace{\rho^{n+1}_{p}}} \left[ -\left(\frac{\partial \delta p}{\partial n}\right)^{n+1}_{f_k} \right],
\label{eq:algorithm-face flux}
\end{equation}
and the updated face velocity $U^{n+1}_{f_{k}}$ satisfies a discrete divergence free condition,
\begin{equation}
    \sum_{f_{k}} U^{n+1}_{f_{k}} A_{f_{k}} = 0.
    \label{eq:divergence_free}
\end{equation}
This updated face flux is saved to be used in the next substep/timestep in Eqs.~\eqref{eq:algorithm-acdi}--\eqref{eq:algorithm-momentum prediction}.

\subsection{Justification for the use of density-weighted velocity in Rhie-Chow interpolation\label{sec:justification of face velocity}}
Recall that the governing equations, Eq.~\eqref{eq:acdi} and Eq.~\eqref{eq:momf}, are in conservative form. Each equation is solved for the phase-field variable $\phi_p$ and the momentum $\Phi_{i,p} = \rho_{p} u_{i,p}$ at cell centers, respectively. Additionally, the density $\rho_p$ is directly obtained from the phase-field variable $\phi_p$ using $\rho_p = \rho_1 \phi_p + \rho_2(1-\phi_p)$, as introduced earlier in Section \ref{sec:consistent mom}. Hereafter, we refer to $\Phi_{i,p}$ and $\rho_p$ as primary variables.  Then, the cell-centered velocity $u_{i,p}$ is defined as the division of $\phi_p$ by $\rho_p$, which may be written as $u_{i,p} \equiv \Phi_{i,p}/\rho_p$. Likewise, the intermediate cell-centered velocity is defined as $u^*_{i,p} = \hat{\Phi}_{i,p}/\rho_{p}$.
In fact, we never solve for the velocity itself directly, it is only a dependent variable that is derived from the two primary variables, $\Phi_{i,p}$ and $\rho_p$. 
Now using these primary variables and expressing 
the velocity gradient at cell faces in terms of these primary variables, we can write:
\begin{eqnarray}
    \left. \frac{\partial u_{i,p}}{\partial n} \right|_f = \left. \frac{\partial}{\partial n} \left(\frac{\rho_{p}u_{i,p}}{\rho_{p}} \right) \right|_f =  \left. \frac{\partial}{\partial n} \left( \frac{\Phi_{i,p}}{\rho_{p}} \right) \right|_f \nonumber\\
    = \left( \left.\frac{\partial \Phi_{i,p}}{\partial n} \right|_f \cellToFace{\rho_{p}} - \cellToFace{\Phi_{i,p}} \left. \frac{\partial \rho_{p}}{\partial n} \right|_f \right) \left( \frac{1}{\cellToFace{\rho_{p}}^2} \right)\\
    =\frac{\rho_{nbr}u_{i,nbr}-\rho_{p}u_{i,p}}{\Delta_f} \frac{1}{\cellToFace{\rho_{p}}} - \frac{\cellToFace{\Phi_{i,p}}}{\cellToFace{\rho_{p}}^2} \frac{\rho_{nbr}-\rho_{p}}{\Delta_f}  = \frac{\rho_{p}\rho_{nbr}}{\cellToFace{\rho_{p}}^2} \frac{u_{i,nbr} - u_{i,p}}{\Delta_f}.
    \label{eq:density weighted velocity gradient}
\end{eqnarray}
Here, we note that the velocity gradient at cell faces, $\left. \frac{\partial u_{i,p}}{\partial n} \right|_f$, computed based on the primary variables will lead to a different expression than from when it is computed directly using the dependent variable.
If we denote the direct calculation of the velocity gradient from the dependent variable $u_{i,p}$ as
\begin{equation}
    \left. \frac{\partial u_{i,p}}{\partial n} \right|_{f,\text{dep.}} = \frac{u_{i,nbr} - u_{i,p}}{\Delta_f},
    \label{eq:velocity gradient dependent variable}
\end{equation}
it is straightforward to notice the two velocity gradients in Eq. \eqref{eq:density weighted velocity gradient}, \eqref{eq:velocity gradient dependent variable} differs by a factor of $\rho_{p}\rho_{nbr}/\cellToFace{\rho_{p}}^2$,
\begin{equation}
    \left. \frac{\partial u_{i,p}}{\partial n} \right|_f = \left. \frac{\partial}{\partial n} \left( \frac{\Phi_{i,p}}{\rho_{p}} \right) \right|_f = \frac{\rho_{p}\rho_{nbr}}{\cellToFace{\rho_{p}}^2} \frac{u_{i,nbr} - u_{i,p}}{\Delta_f} = \frac{\rho_{p}\rho_{nbr}}{\cellToFace{\rho_{p}}^2} \left. \frac{\partial u_{i,p}}{\partial n} \right|_{f,\text{dep.}}.
    \label{eq:velocity gradient primary variables}
\end{equation}
Since the geometric mean is always less than or equal to the arithmetic mean, the velocity gradient using the primary variables in Eq.~\eqref{eq:velocity gradient primary variables} is always less than
the velocity gradient using the $u_{i,p}$ directly in Eq.~\eqref{eq:velocity gradient dependent variable}, and is identical for the unity-density ratio. Consequently, the face velocity $U_f$ is not a simple average of two neighboring cell velocities,
\begin{equation}
    U_f = u_{i,p} + \left. \frac{\partial u_{i,p}}{\partial n} \right|_f \frac{\Delta_f}{2} \neq \frac{u_{i,p} + u_{i,nbr}}{2}.
    \label{eq:face velocity}
\end{equation}

\begin{figure}
    \centering
    \includegraphics[width=0.5\textwidth]{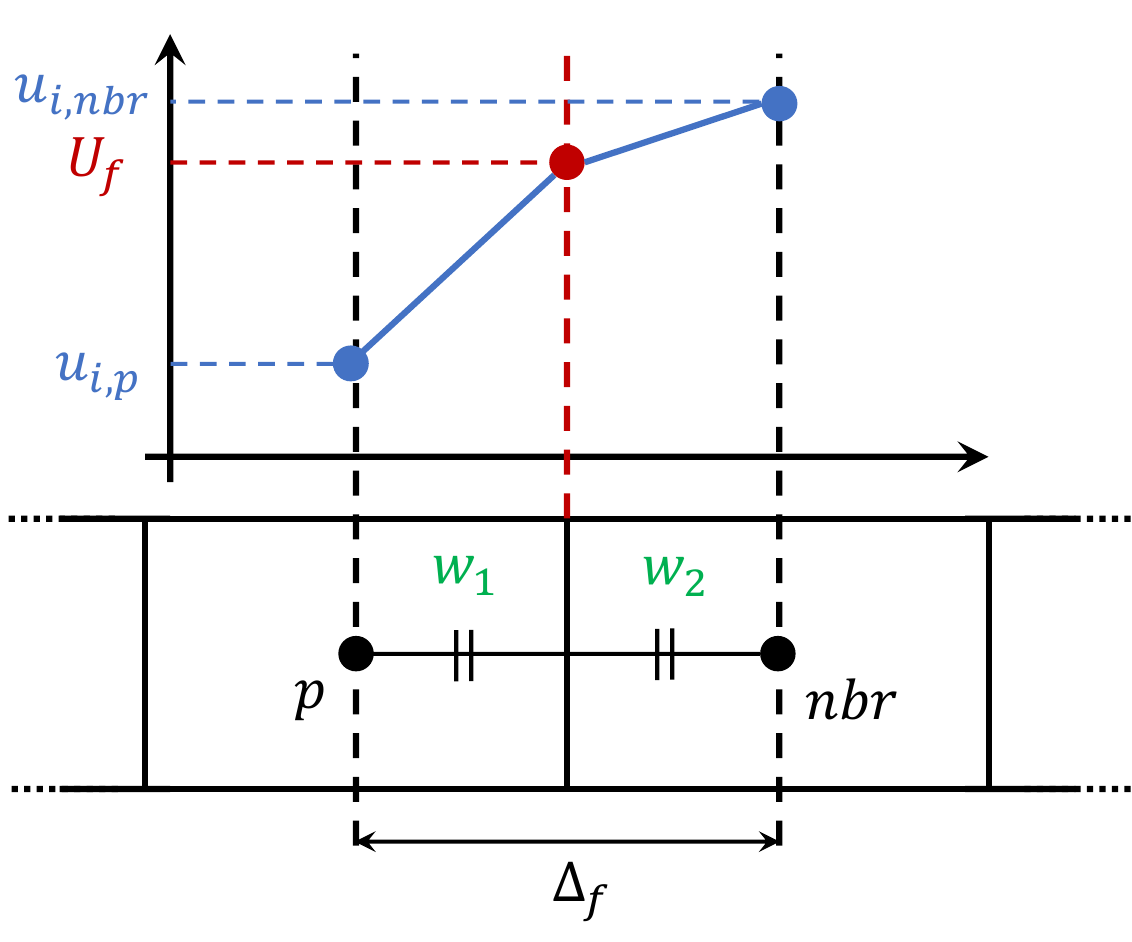}
    \caption{A simplified 1D schematic of a cell face and two neighbouring cells to compute the face velocity $U_f$.}
    \label{fig:face velocity schematic}
\end{figure}

This result implies some sort of weighting is required to compute the face velocity $U_f$ from the dependent variable $u_{i,p}$. Then, our task at hand is to compute $U_f$ by finding the weights $w_1$ and $w_2$, as shown in Figure~\ref{fig:face velocity schematic}, which will give $U_f$ that can be used in pressure projection step (Eq. \eqref{eq:algorithm-pressure Poisson}) and the calculation of updated face flux (Eq. \eqref{eq:algorithm-face flux}).
To proceed further, we assume $U_f$ as
\begin{equation}
    U_f = w_1 u_{i,p} + w_2 u_{i,nbr},
    \label{eq:weighted face velocity}
\end{equation}
where $w_1$ and $w_2$ satisfy $\sum w_i = 1$.
In addition, the velocity gradient in Eq.~\eqref{eq:velocity gradient primary variables} is computed by the velocity gradients from each half-segment ($\Delta_f/2$) in Figure~\ref{fig:face velocity schematic} with the same weights as
\begin{equation}
    \left. \frac{\partial u_{i,p}}{\partial n} \right|_f = w_1 \frac{U_f - u_{i,p}}{\Delta_f/2} + w_2 \frac{u_{i,nbr} - U_f}{\Delta_f/2}.
    \label{eq:weighted velocity gradient}
\end{equation}
Solving Eqs.~\eqref{eq:weighted face velocity}--\eqref{eq:weighted velocity gradient} for the weights $w_i$ yield two sets of solutions,
\begin{equation}
    w_1 = \frac{\rho_{p}}{\rho_{p} + \rho_{nbr}},~w_2 = \frac{\rho_{nbr}}{\rho_{p} + \rho_{nbr}} \quad \text{or} \quad w_1 = \frac{\rho_{nbr}}{\rho_{p} + \rho_{nbr}},~w_2 = \frac{\rho_{p}}{\rho_{p} + \rho_{nbr}}.
    \label{eq:weight solution}
\end{equation}
As previously mentioned in deriving Eq.~\eqref{eq:algorithm-update face flux}, our choice of weights is the first set because $w_1$ corresponds to the quantity associated with cell $p$ and $w_2$ corresponds to the quantity associated with the neighboring cell $nbr$ which will lead to the consistent cell-to-face interpolation.

Finally, substitution of the first set solution in Eq.~\eqref{eq:weight solution} for the weights to  Eq.~\eqref{eq:weighted face velocity} yields the face velocity $U_f$ as the density-weighted velocity given by
\begin{equation}
    U_f = \frac{\cellToFace{\Phi_{i,p}}}{\cellToFace{\rho_{p}}} = \frac{\rho_{p}u_{i,p}+\rho_{nbr}u_{i,nbr}}{\rho_{p}+\rho_{nbr}}.
    \label{eq:final weighted face velocity}
\end{equation}
This way of computing the face velocity $U_f$ can be interpreted as the division between the face-interpolated primary values, $\cellToFace{\Phi_{i,p}}$ over $\cellToFace{\rho_p}$ evaluated at face, which is consistent with how the dependent variable $\vec{u}$ is computed at cell centers. In fact, it turns out that other choices of $U_f$, such as a simple arithmetic average $(u_{i,p}+u_{i,nbr})/2$ that is typically used in the literature or the resultant of substituting the second set of weights in Eq.~\eqref{eq:weight solution}, $(\rho_{p}u_{i,nbr}+\rho_{nbr}u_{i,p})/(\rho_{p}+\rho_{nbr})$, will lead to failure of the simulation with unbounded growth of total kinetic energy during long integrations, making the method non-robust (see, discussion in Section \ref{sec:energy-stable collocated grid formulation}).

In summary, we propose to compute the intermediate face velocity $\hat{U}_f$ as a density-weighted velocity during the Rhie-Chow interpolation step in Eq. \eqref{eq:algorithm-update face flux}, which will then be used for pressure projection (Eq. \eqref{eq:algorithm-pressure Poisson}) and the computation of divergence free velocity (Eq. \eqref{eq:algorithm-face flux}),
mainly based on two reasons. 
First, the discrete velocity gradients are evaluated using primary variables, which will lead to consistent computation of velocity both at cell faces and cell centers.
Second, the face velocity follows the consistent cell-to-face interpolation, e.g., the first set of the weights in Eq.~\eqref{eq:weight solution}, as the last term in Eq.~\eqref{eq:algorithm-update face flux} is discretized.
Going back to the evaluation of the first term of Eq.~\eqref{eq:algorithm-update face flux}, and as will be shown in Eq.~\eqref{eq:energy budget: error analysis 1} below, the density-weighted intermediate face velocity is employed instead of a simple average of neighboring cell center velocities (that is typically what is used in the literature) to compute the discrete divergence of cell-centered velocity $u_{i,p}$.

\section{Discrete energy conservation\label{sec:discrete energy conservation}}
It is straightforward to show that the convective terms and the interface-regularization terms do not contribute to global discrete kinetic energy due to the skew-symmetric form of the flux-split operators \citep{jain2022kinetic}. However, the discretization used for the pressure gradient term has to be carefully chosen. 

\subsection{Staggered grid formulation}

In a staggered grid formulation, the pressure gradient contribution in the momentum equation can be written as
$$
\frac{\partial \left( \langle \rho_{p} \rangle_{p\rightarrow f} v_{f} \right) }{\partial t}= - \left(\frac{p_{nbr}-p_{p}}{d_f}\right) + \cdots
$$
where $v_{f}$ is the face projected velocity between $p$ and $nbr$ cells, and $d_{f}=|\mathbf{x_p}-\mathbf{x_{nbr}} |$. 
Then the contribution of the first term in the above equation to the kinetic energy (defined on the cell faces) can be computed as
$$
v_{f} \frac{\partial \left( \langle \rho_{p} \rangle_{p\rightarrow f} v_{f} \right) }{\partial t} A_{f} d_{f}=-v_{f}\left(p_{nbr}-p_{p}\right) A_{f}.
$$
Summing over all the faces in the domain yields,
$$
\sum_{f} \frac{\partial}{\partial t}\left( \langle \rho_{p} \rangle_{p\rightarrow f} \frac{v_f^2}{2} A_f d_f\right)
=
-\sum_{f} v_{f}\left(p_{nbr}-p_{p}\right) A_{f},
$$
which reduces to 
$$
\sum_{f} 
\frac{\partial}{\partial t}\left( \langle \rho_{p} \rangle_{p\rightarrow f} \frac{v_f^2}{2} A_f d_f\right)
=
\sum_{p} p_{p} \underbrace{ {{\sum_{\text{interior} f}} v_{f} A_{f}} }_{=0} -\sum_{\text{boundary} f} v_{f} p_{f} A_{f}.
$$
Therefore, the pressure gradient term doesn't contribute to the global discrete kinetic energy in a fully staggered grid formulation. 

\subsection{Energy-stable collocated grid formulation\label{sec:energy-stable collocated grid formulation}}

In this work, a density-weighted velocity interpolation is proposed to compute the intermediate velocity on faces in Eq. \eqref{eq:algorithm-update face flux}. In this section, we will show that this step is crucial in obtaining an energy-stable method.

\subsubsection{Global kinetic energy in continuous form}

Contracting the momentum equation in Eq.~\eqref{eq:momf} with the velocity vector $u_{i,p}$, as
\begin{equation}
    u_{i,p} \left( \frac{\partial \rho u_i}{\partial t} + \frac{\partial \rho u_i u_j}{\partial x_j} =  -\frac{\partial p}{\partial x_i} + \frac{\partial u_i f_j}{\partial x_j} \right).
    \label{eq:energy budget derivation: continuous form 1}
\end{equation}
Here, we have omitted the viscous term and surface tension forces for brevity. With algebraic manipulations, it can be written as
\begin{eqnarray}
    \frac{\partial}{\partial t} \left( \frac{1}{2} \rho_p u_{i,p}u_{i,p} \right) + \frac{\partial}{\partial x_j} \left( \frac{1}{2} \rho_p u_{i,p}u_{i,p}u_{j,p} \right) + \frac{1}{2} u_{i,p}u_{i,p} \left( \frac{\partial \rho_p}{\partial t} + \frac{\partial \rho_p u_{i,p}}{\partial x_j} \right) = \nonumber \\
    u_{i,p} \frac{\partial}{\partial x_j} \left( u_{i,p} f_{j,p} \right) - \frac{\partial p_p u_{i,p}}{\partial x_i} + p_{p} \frac{\partial u_{i,p}}{\partial x_i}.
    \label{eq:energy budget derivation: continuous form 2}
\end{eqnarray}
Substitution of the implied continuity equation in Eq.~\eqref{eq:mod_continuity} further reduces Eq.~\eqref{eq:energy budget derivation: continuous form 2} to
\begin{equation}
    \frac{\partial}{\partial t} \left( \frac{1}{2} \rho_p u_{i,p}u_{i,p} \right) + \frac{\partial}{\partial x_j} \left( \frac{1}{2} \rho_p u_{i,p}u_{i,p}u_{j,p} \right) -  \frac{\partial}{\partial x_j} \left( \frac{1}{2} u_{i,p} u_{i,p} f_{j,p} \right) = - \frac{\partial p_p u_{i,p}}{\partial x_i} + p_{p} \frac{\partial u_{i,p}}{\partial x_i}.
    \label{eq:energy budget derivation: continuous form 3}
\end{equation}
All the terms in Eq.~\eqref{eq:energy budget derivation: continuous form 3} are written in conservative form except the last term. 

\subsubsection{Global kinetic energy in discrete form}

Similar to the continuous form above, the kinetic energy conservation property can be investigated in a discrete sense by taking a discrete dot product of the velocity with the discrete momentum equation. Here we look at the terms in the momentum equation individually. We begin with the unsteady term, which can be approximated as
\begin{eqnarray}
    \left( \frac{u^{n+1}_{i,p} + u^n_{i,p}}{2} \right) \left( \frac{\rho^{n+1}_p u^{n+1}_{i,p} - \rho^{n}_p u^{n}_{i,p}}{\Delta t} \right) \nonumber \\
    = \frac{1}{\Delta t} \left( \frac{1}{2} \rho^{n+1}_p u^{n+1}_{i,p} u^{n+1}_{i,p} - \frac{1}{2} \rho^{n}_p u^{n}_{i,p} u^{n}_{i,p} \right) + \frac{1}{2} u^{n+1}_{i,p} u^{n}_{i,p} \left( \frac{\rho^{n+1}_p - \rho^{n}_p}{\Delta t} \right).
    \label{eq:energy budget derivation: discrete form 1}
\end{eqnarray}
The last term in Eq.~\eqref{eq:energy budget derivation: discrete form 1} can be further expanded using the continuity equation as
\begin{eqnarray}
    \frac{1}{\Delta t} \left( \frac{1}{2} \rho^{n+1}_p u^{n+1}_{i,p} u^{n+1}_{i,p} - \frac{1}{2} \rho^{n}_p u^{n+1}_{i,p} u^{n}_{i,p} \right) + \frac{1}{2} u^{n}_{i,p} u^{n}_{i,p} \left( -\sum_{f_k} \frac{\rho^{n}_{p} + \rho^{n}_{nbr}}{2} U^{n}_{f} A_{f_k} + \sum_{f_k} f^{n}_{f_k} A_{f_k}\right).
    \label{eq:energy budget derivation: discrete form 2}
\end{eqnarray}
Similarly, the dot product with the interface-regularization term in the momentum equation will yield
\begin{eqnarray}
    - u^{n}_{i,p} \sum_{f_k} \left( \frac{u^{n}_{i,p} + u^{n}_{i,nbr}}{2} \right) f^n_{f_k} A_{f_k} = -\sum_{f_k} \left( \frac{u^{n}_{i,p} u^{n}_{i,p} + u^{n}_{i,p} u^{n}_{i,nbr}}{2} \right) f^n_{f_k} A_{f_k} \nonumber \\
    = -\sum_{f_k} \left( \frac{u^{n}_{i,p} u^{n}_{i,nbr}}{2} \right) f^n_f A_{f_k} - \frac{u^{n}_{i,p} u^{n}_{i,p}}{2} \sum_{f_k} f^n_{f_k} A_{f_k}.
    \label{eq:energy budget derivation: discrete form 3}
\end{eqnarray}
Summation of Eq.~\eqref{eq:energy budget derivation: discrete form 2} and Eq.~\eqref{eq:energy budget derivation: discrete form 3} cancels out the last terms of each equation, respectively. So far, the only term in non-conservative form is the third term in Eq.~\eqref{eq:energy budget derivation: discrete form 2}. This remaining term can be further manipulated along with the convection term as
\begin{eqnarray}
    u^{n}_{i,p} \sum_{f_k} \left( \frac{\rho_p + \rho_{nbr}}{2} \right) \left( \frac{u_{i,p} + u_{i,nbr}}{2} \right) U^n_{f_k} A_{f_k} - \frac{1}{2} u^{n}_{i,p} u^{n}_{i,p} \left( \sum_{f_k} \frac{\rho^{n}_{p} + \rho^{n}_{nbr}}{2} U^{n}_{f_k} A_{f_k} \right) \nonumber \\
    = \sum_{f_k} \left( \frac{\rho^{n}_{p} + \rho^{n}_{nbr}}{2} \right) u^{n}_{i,p} u^{n}_{i,nbr} U^{n}_{f_k} A_{f_k},
    \label{eq:energy budget derivation: discrete form 4}
\end{eqnarray}
and together they can be written in a conservative form. Consequently, combining the formulations from Eqs. \eqref{eq:energy budget derivation: discrete form 2}-\eqref{eq:energy budget derivation: discrete form 4} and rearranging the equation with the remaining pressure-related term at its RHS, we obtain a form of the discrete kinetic energy equation as
\begin{eqnarray}
    \frac{1}{\Delta t} \left( \frac{1}{2} \rho^{n+1}_p u^{n+1}_{i,p} u^{n+1}_{i,p} - \frac{1}{2} \rho^{n}_p u^{n}_{i,p} u^{n}_{i,p} \right)
    +\sum_{f_k} \left( \frac{\rho^{n}_{p} + \rho^{n}_{nbr}}{2} \right) u^{n}_{i,p} u^{n}_{i,nbr} U^{n}_{f_k} A_{f_k}
    -\sum_{f_k} \left( \frac{u^{n}_{i,p} u^{n}_{i,nbr}}{2} \right) f^n_f A_{f_k} \nonumber \\
    = -u_{i,p} \sum_{f_k} \frac{p^{n}_{i,p} + p^{n}_{i,nbr}}{2} A_{f_k}
    = -\sum_{f_k} \frac{p_p u_{i,nbr} + p_{nbr} u_{i,p}}{2} A_{f_k} + p^{n}_{i,p} \sum_{f_k} U^n_{f_k} A_{f_k}.
    \label{eq:energy budget derivation: discrete form 5}
\end{eqnarray}

As pointed out by~\citet{ham2004}, the magnitude of the discrete divergence, which is the last term in Eq.~\eqref{eq:energy budget derivation: discrete form 5}, must be minimized to keep the discrete kinetic energy conservation as much as possible.
As proposed in Section~\ref{sec:projection method algorithm}, the discrete divergence is written as
\begin{eqnarray}
    \sum_{f_k} U^{n}_{f_k} A_{f_k} = \sum_{f_k} \left[ \frac{\rho_p u_{i,p} + \rho_{nbr} u_{i,nbr}}{\rho_{p}+\rho_{nbr}} \right]^{n} A_{f_k} \nonumber \\
    = \sum_{f_k} \frac{\cellToFace{\hat{\Phi}_{i,p}^{}}}{\cellToFace{\rho_{p}^{}}} n_{i,f_k} A_{f_k} - \sum_{f_k} \frac{\Delta t}{\cellToFace{\rho_{p}^{n}}} \frac{1}{2} \left[ \left(\frac{\partial p}{\partial x_i}\right)_{p} + \left(\frac{\partial p}{\partial x_i}\right)_{nbr}\right]^{n} n_{i,f_k} A_{f_k}
    \nonumber \\
    = \underbrace{ \sum_{f_k} \left( \frac{\cellToFace{\hat{\Phi}_{i,p}}}{\cellToFace{\rho_{p}^{n+1}}} -\frac{\Delta t}{\cellToFace{\rho_{p}^{n}}} \left(\frac{\partial p}{\partial n}\right)^{n}_{f_k} \right) A_{f_k} }_{=0~\textnormal{due to Eq.}~\eqref{eq:divergence_free}}\nonumber \\
    + \sum_{f_k} \left( \frac{\Delta t}{\cellToFace{\rho_{p}^{n}}} \left(\frac{\partial p}{\partial n}\right)^{n}_{f_k} 
    - \frac{\Delta t}{\cellToFace{\rho_{p}^{n}}} \frac{1}{2} \left[ \left(\frac{\partial p}{\partial x_i}\right)_{p} + \left(\frac{\partial p}{\partial x_i}\right)_{nbr}\right]^{n} n_{i,f_k}\right) A_{f_k}.
    \label{eq:energy budget: error analysis 1}
\end{eqnarray}

Note we have used Eqs.~\eqref{eq:algorithm-update face flux reduced form}--\eqref{eq:algorithm-update face flux}.
With Taylor-series expansions, 
\begin{eqnarray}
    \left(\frac{\partial p}{\partial x_i}\right)_p n_{i,f} = \left( \frac{\partial p}{\partial n} \right)_{f} - \left( \frac{\partial^2 p}{\partial x_i^2} \right)_{p} n_{i,f} \left( \frac{\Delta_f}{2} \right) + \dots \nonumber \\
    \left(\frac{\partial p}{\partial x_i}\right)_{nbr} n_{i,f} = \left( \frac{\partial p}{\partial n} \right)_{f} + \left( \frac{\partial^2 p}{\partial x_i^2} \right)_{nbr} n_{i,f} \left( \frac{\Delta_f}{2} \right) + \dots,
    \label{eq:energy budget: error analysis 2}
\end{eqnarray}
Eq.~\eqref{eq:energy budget: error analysis 1} is further approximated as
\begin{eqnarray}
    -\sum_{f_k} \frac{\Delta t}{2\rho_f} \frac{\Delta_f}{2} \left\{ -\left( \frac{\partial^2 p}{\partial x_i^2} \right)_{p} + \left( \frac{\partial^2 p}{\partial x_i^2} \right)_{nbr} \right\} n_{i,f} A_{f_k} \nonumber \\
    \approx -\Delta t \sum_{f_k} \frac{\Delta_f^2}{4 \rho_f} \left( \frac{\partial^3 p}{\partial n^3} \right)_{f_k} A_{f_k}
    \approx -\frac{\Delta t\Delta_f^2}{4} \frac{1}{\rho_p} \left( \frac{\partial^4 p}{\partial x_i^4} \right)_{p} V_{cv}.
    \label{eq:energy budget: error analysis 3}
\end{eqnarray}
Then, it is straightforward to show that the last resultant term in Eq.~\eqref{eq:energy budget: error analysis 3} is strictly dissipative in nature. The discrete equivalent modified equation of Eq.~\eqref{eq:energy budget derivation: continuous form 3} can then be written as
\begin{eqnarray}
    \frac{\partial}{\partial t} \left( \frac{1}{2} \rho_p u_{i,p}u_{i,p} \right) + \frac{\partial}{\partial x_j} \left( \dots \right) 
    = - \left( \frac{\Delta t\Delta_f^2}{4 \rho_p} \right) p \left( \frac{\partial^4 p}{\partial x_i^4} \right)_{p} \nonumber \\
    = - \left( \frac{\Delta t\Delta_f^2}{4 \rho_p} \right) \left( \frac{\partial}{\partial x_j} \left( p \frac{\partial^3 p}{\partial x^3_j} - \frac{\partial p}{\partial x_j} \frac{\partial^2 p}{\partial x^2_j}\right) + \left( \frac{\partial^2 p}{\partial x_j^2}\right)^2 \right),
    \label{eq:energy budget: error analysis 4}
\end{eqnarray}
where $(\dots)$ contains all the other terms in conservative form.
The first term is in divergence form, however, the squared form of the last term is strictly positive and hence causes the dissipation of the total kinetic energy. Hence, this formulation is energy stable.
The dissipation error is investigated for a droplet-laden HIT field in Section~\ref{sec:result-HIT} below, and the results are provided in Figure~\ref{fig:hit kinetic energy error analysis}.

\subsection{Gradient reconstruction scheme\label{sec:gradient reconstruction scheme}}
As was shown in Section \ref{sec:energy-stable collocated grid formulation}, fractional-step method (Section~\ref{sec:projection method algorithm}) has a dissipative error in the transport of global kinetic energy, which is originated from pressure gradient term in a collocated grid framework. To conserve global kinetic energy as much as possible, the magnitude of the last term in Eq.~\eqref{eq:energy budget derivation: discrete form 5} evaluated at cell center (see, the last term in Eq.~\eqref{eq:energy budget: error analysis 4}) should be minimized. This minimization of the error can be achieved at the face-to-cell interpolation step, when evaluating the last term in Eq.~\eqref{eq:algorithm-cell center velocity}, using the least square method. Note that the external forces such as surface tension and gravity are included in the final form of Eq.~\eqref{eq:algorithm-cell center velocity}. In the derivation of the final result in Eq.~\eqref{eq:energy budget: error analysis 1}, including external forces leads to 
\begin{eqnarray}
\sum_{f_k}\left\{ \frac{\Delta t}{\langle \rho^{n+1}_{p} \rangle_{p\rightarrow f_k}} \left[\left(\frac{\partial p}{\partial n}\right)^{n+1}_{f_k} - \left(F_{j,f_k} n_{j,f_{k}}\right)^{n+1}\right]
\right\} A_{f_k} \nonumber \\
- \frac{\Delta t}{2} \sum_{f_k}\left[ 
 \frac{1}{ \rho^{n+1}_{p}}\left(\frac{\partial p}{\partial x_i}\right)^{n+1}_{p} 
+ \frac{1}{ \rho^{n+1}_{nbr_k}}\left(\frac{\partial p}{\partial x_i}\right)^{n+1}_{nbr_k} \right] n_{i,f} A_{f_k}
+ \Delta t  \sum_{f_k}\left( \frac{F_{i,p}^{n+1} + F_{i,nbr_k}^{n+1} }{2} \right) n_{i,f} A_{f_k},
\label{eq:gradient reconstruction:error minimization 1}
\end{eqnarray}
Then, the remaining task at hand is to minimize the difference between $\frac{1}{2}\left((\frac{\partial p}{\partial x_i})_{p}+(\frac{\partial p}{\partial x_i})_{nbr}\right) N_i$ across each face and $\partial p / \partial n$ for pressure gradient terms and the same concept is applied to external force.
However, since quantities evaluated at cell center, such as $\partial p / \partial x_i$, are located at the volumes, while surface-evaluated quantity, such as $\partial p / \partial n$, is located at the faces, Eq.~\eqref{eq:gradient reconstruction:error minimization 1} cannot be imposed exactly. Instead, our approach is to minimize
\begin{align}
\sum_{f_k} \left\{ \left[ 
 \frac{1}{ \rho^{n+1}_{p}}\left(\frac{\partial p}{\partial x_i}\right)^{n+1}_{p} - F_{i,p}^{n+1}  \right] n_{i,f}
- \frac{1}{\langle \rho^{n+1}_{p} \rangle_{p\rightarrow f_k}} \left[\left(\frac{\partial p}{\partial n}\right)^{n+1}_{f_k} - \left(F_{j,f_k} n_{j,f_{k}}\right)^{n+1}\right]
\right\} A_{f_k},
\label{eq:gradient reconstruction:error minimization 2}
\end{align}
in a least squares sense. This least squares procedure is defined with an operator $\lsOp{\cdot}$, which maps the values evaluated at cell faces, for example $\partial \theta/\partial n$, to values at cell center $\partial \theta/\partial x_i$, by solving
\begin{equation}
    \sum_{f_k} \mathbf{n}_f \otimes \mathbf{n}_f \nabla \theta A_{f_k} = \sum_{f_k} \mathbf{n}_f \mathbf{n}^{\textnormal{T}}_f \nabla \theta A_{f_k} = \sum_{f_k} \nabla \theta \cdot \mathbf{n}_f A_{f_k},
    \label{eq:gradient reconstruction:error minimization 3}
\end{equation}
which is equivalent to
\begin{equation}
\setlength{\arraycolsep}{4pt}
\renewcommand{\arraystretch}{1.3}
\sum_{f_k}
\left[
\begin{array}{ccc}
n_{xpj}n_{xpj} &n_{xpj}n_{ypj} &n_{xpj}n_{zpj}\\
n_{xpj}n_{ypj} &n_{ypj}n_{ypj} &n_{ypj}n_{zpj}\\
n_{xpj}n_{zpj} &n_{ypj}n_{zpj} &n_{zpj}n_{zpj}
\end{array}
\right]
\left[
\begin{array}{c}
\frac{\partial \theta_p}{\partial x}\\
\frac{\partial \theta_p}{\partial y}\\
\frac{\partial \theta_p}{\partial z}
\end{array}
\right]
A_{f_k}
=
\sum_{f_k}
\left[
\begin{array}{c}
\left. \frac{\partial \theta_p}{\partial x_i}\right|_{f} \cdot n_{ij}
\end{array}
\right]
\left[
\begin{array}{c}
n_{xpj}\\
n_{ypj}\\
n_{zpj}
\end{array}
\right]
A_{f_k}
.
\label{eq:gradient reconstruction:error minimization 4}
\end{equation}

As an alternative to the least squares procedure, results using the Green Gauss procedure are presented in~\ref{sec:app:green gauss}.

\section{Results \label{sec:results}}
In this section, we present the simulation results of the accurate conservative phase-field/diffuse-interface (ACDI) method on Voronoi unstructured grids. This phase-field method is solved in conjunction with a novel energy-stable fractional-step method for two-phase flows on collocated grids proposed in this work. We first present results using canonical test cases in addition to new test cases with grid transition that we have designed to systematically evaluate interface-capturing methods for simulation of two-phase flows on unstructured grids. Advection of a single drop with initially uniform velocity field, a drop in a vortex, and a droplet-laden homogeneous isotropic turbulence (HIT) are used as the test cases. Using these test cases, we investigate the accuracy of the method, boundedness of the PF variable $\phi$, total volume (mass) conservation, and the robustness of the method by tracking the evolution of global kinetic energy change. We define the shape error as,
\begin{equation}
    \mathcal{E}=\sum_{cv}|\phi_{cv}-\phi_{0,cv}|dV^{\frac{2}{3}}_{cv},
    \label{eq:error metric}
\end{equation}
which is used as a quantitative metric to evaluate the accuracy of the resulting interface shape, 
where $\phi_0$ denotes the initial PF field at time $t=0$.
For all the simulations, $\Gamma$ and $\epsilon^*$ are chosen as $\Gamma = 1.0 \times |u|_{max}$ and $\epsilon^*= \epsilon/\Delta_f = 0.6$, where $|u|_{max}$ is the maximum velocity in the flow field and $\Delta_f$ is a local characteristic length as introduced in Section \ref{sec:ACDI-discrete}.
In addition to the local characteristic length, we also provide the shape errors with a constant $\Delta_f$ for the drop-advection and drop-in-a-vortex test cases, where the simulations were rerun with $\Delta_f=\Delta_{const}$. The time-step size is set to $\Delta t = 0.001$ unless otherwise specified.

At the end, more complex test cases are presented in Section~\ref{sec:complex two phase flow apps}, demonstrating the verification and validation efforts of the proposed method using a damped surface wave, a liquid jet with optimized disturbance, and a liquid jet atomization from a non-cavitating Spray A nozzle from the engine combustion network (ECN).

\subsection{Drop advection}
For the advection test case, we consider a two-dimensional unit square domain $[0.0,1.0]\times [0.0,1.0]$ with a drop of radius $r=0.15$ initially located at the center $(x,y)=(0.5,0.5)$. 
The initial velocity field in the domain is $(u,v)=(1.0,0.0)$ unless otherwise specified, and the simulations were run until $t = 4$, which is $4$ periods of drop advection in the $x$ direction. We note that the ACDI model in Eq.~\eqref{eq:acdi} is coupled with the consistent momentum equation in Eq.~\eqref{eq:momf}. Multiple simulations with different grid configurations were used to evaluate the effect of grid transition on the accuracy of the interface-capturing method. Details of the different grids along with the setup used for each simulation are described below:
\begin{enumerate}
    \item Case 1: A uniform HCP grid of $128\times128$ cells.
    \item Case 2: A base HCP grid of $64\times64$ cells with refinement in the region of $y\in[0.3,0.7]$ by a factor of $4$. Additionally, there are $5$ cells with a refinement factor of $2$ in the grid transition region where there is a nested refinement. The drop is always fully inside the refined region, as it advects, in this case.
    \item Case 3: A base HCP grid of $64\times64$ cells with refinement in the region of $x\in[0.48,0.52]$ by a factor of $4$. Cell-count in the nested refinement region, where the cells are refined by a factor of $2$, is $4$. The drop advection direction is parallel to the gradient of grid transition, so the drop passes through the grid transition region twice for each period of advection.
    \item Case 4: A base HCP grid of $64\times64$ cells with refinement in the region of $y\in[0.48,0.52]$ by a factor of $4$. Cell-count in the nested refinement region, where the cells are refined by a factor of $2$, is $4$. The drop advection direction is perpendicular to the gradient of grid transition, so the drop is always in the grid transition region.
    \item Case 5: A uniform HCP grid of $128\times128$ cells, but the drop advects diagonally.
\end{enumerate}
To fully evaluate the accuracy of the method in various situations, interface advection through (Case 3) and along (Case 4) grid transitions and domains with partially refined regions were studied. These test cases are important because the accuracy of a phase-field method is dependent on the truncation error in the numerical scheme used, which is locally a function of the grid topology.

The initial velocity field for each simulation and the corresponding shape error computed using Eq.~\eqref{eq:error metric} are given in Table~\ref{tab:shape error:drop advection}, where $\Delta_\text{const}$ represents a constant characteristic length, which is identical to the distance between cell centers in a uniform grid.
$\mathcal{E}(\Delta_f)$ and $\mathcal{E}(\Delta_{\text{const}})$ are shape errors when local characteristic length $\Delta_f = \Delta_{\text{local}}$ and a constant length $\Delta_f = \Delta_{\text{const}}$ are used, respectively.

\begin{table}
\begin{center}
\setlength{\tabcolsep}{12pt}
\begin{tabular}{c c c c c c}
\hline
Case ID & \makecell{Initial velocity\\field $(u,v)$} & $\Delta t$ & $\Delta_{\text{const}}$  & $\mathcal{E}(\Delta_{\text{local}})$ & $\mathcal{E}(\Delta_{\text{const}})$\\
\hline
\hline
Case 1 & ($1.0$, $0.0$) & $1.0\times10^{-3}$ & $1/128$ & $8.329\times10^{-4}$ & $2.330\times10^{-3}$\\
Case 2 & ($1.0$, $0.0$) & $1.0\times10^{-3}$ & $1/256$ & $3.011\times10^{-4}$ & $1.412\times10^{-3}$\\
Case 3 & ($1.0$, $0.0$) & $4.0\times10^{-4}$ & $1/64$  & $2.376\times10^{-3}$ & $5.343\times10^{-3}$\\
Case 4 & ($1.0$, $0.0$) & $4.0\times10^{-4}$ & $1/64$  & $5.029\times10^{-3}$ & $4.174\times10^{-3}$\\
Case 5 & ($1.0$, $1.0$) & $1.0\times10^{-3}$ & $1/128$ & $1.117\times10^{-3}$ & $3.559\times10^{-3}$\\
\hline
\end{tabular}
\end{center}
\caption{Parameters used for droplet advection cases and shape errors for simulations with constant $\mathcal{E}(\Delta_{\text{const}})$ and local characteristic length $\mathcal{E}(\Delta_{\text{local}})$.}
\label{tab:shape error:drop advection}
\end{table}

\begin{figure}
    \centering
    \includegraphics[width=\textwidth]{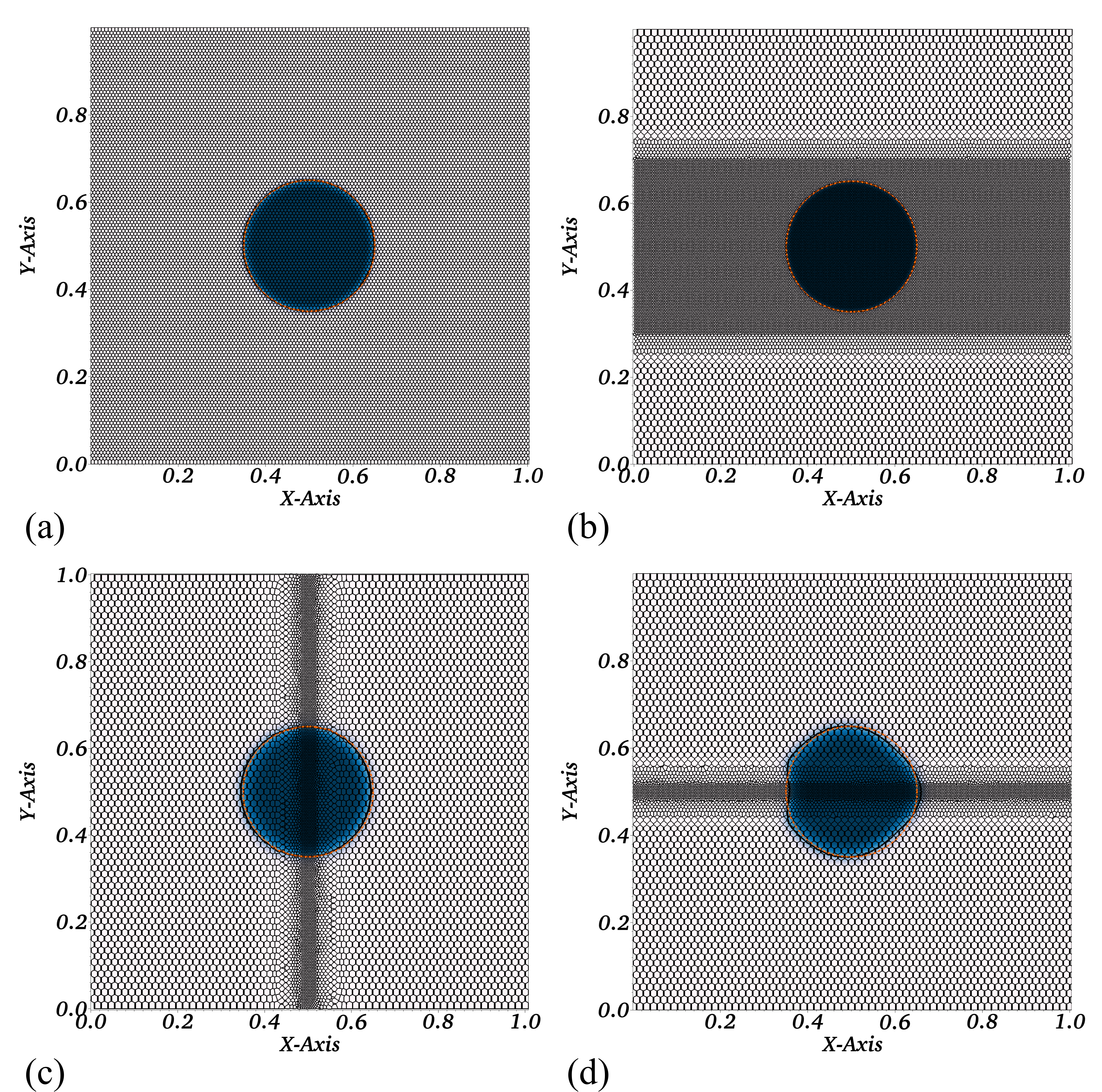}
    \caption{Snapshots of $\phi$ field for the advection test cases at the final time of $t=4$ for (a) case 1, (b) case 2, (c) case 3, and (d) case 4. The orange and black solid lines represent the initial and final drop (isocontour of $\phi=0.5$), respectively. Blue and white cell color indicate $\phi$ values of unity and zero, respectively. Black solid lines represent the mesh.}
    \label{fig:fig1}
\end{figure}

The simulation snapshots from the droplet advection case are shown in Figure~\ref{fig:fig1} at both the initial $t=0$ and the final time $t=4$. We confirmed that the total volume $\sum \phi_{cv} \textnormal{d}V_{cv}$ was conserved and the boundedness of $\phi$ was preserved for all cases. For cases $3$ and $4$, the shape error accumulates faster compared to other cases as the droplet encounters regions of grid transition. The shape error for case $3$ is better than case $4$ because the drop encounters the grid transition only twice per period of advection, as opposed to advection always being along the grid transition. This suggests that the present method results in accurate simulations of two-phase flows in realistic simulations of engineering interest, where one would typically use complex meshes with grid transitions. For cases $1$, $2$, and case $5$ (not shown in Figure~\ref{fig:fig1}) where the drop does not encounter grid transition regions, the shape error is sufficiently low, and the initial drop shape is well maintained.

In Table~\ref{tab:shape error:drop advection}, the shape errors with two different characteristic lenghts, one with a constant characteristic length $\mathcal{E}(\Delta_{\text{const}})$ and the other with a local characteristic length $\mathcal{E}(\Delta_{\text{local}})$, show comparable shape errors, showing the maximum error difference between the two sets of shape errors less than $0.4\%$, for all cases. Additionally, we note that the use of local characteristic length allows larger time step more than twice as much as the ones with constant characteristic length. 

\subsection{Drop in a vortex}
In this section, we evaluate the method using a drop-in-a-vortex case where the drop undergoes shearing deformation. We consider a two-dimensional unit square domain $[0.0,1.0]\times [0.0,1.0]$ with a uniform HCP mesh of size $128 \times 128$.
Initially, a drop of radius $r=0.15$ is placed at $(x,y) = (0.5, 0.75)$. Here, the ACDI method (Eq.~\eqref{eq:acdi}) is decoupled from the momentum equation (Eq.~\eqref{eq:momf}), and instead a velocity field of
\begin{eqnarray}
    u = -\sin^2(\pi x)\sin(2\pi y)\cos\left(\frac{\pi t}{T}\right),\\
    v = \sin(2\pi x)\sin^2(\pi y)\cos\left(\frac{\pi t}{T}\right),
\end{eqnarray}
is prescribed at every time step. Here, the period $T$ is set as $T = 4$. 

\begin{table}
\begin{center}
\setlength{\tabcolsep}{12pt}
\begin{tabular}{c c c c c}
\hline
Case ID & $\Delta_{\text{const}}$  & $\mathcal{E}(\Delta_{\text{local}})$ & $\mathcal{E}(\Delta_{\text{const}})$ & $\mathcal{E}(\Delta_{32})$\\
\hline
\hline
Case 6 & $1/32$  & $3.098\times10^{-2}$ & $3.347\times10^{-2}$ & $3.347\times10^{-2}$\\
Case 7 & $1/64$  & $8.602\times10^{-3}$ & $8.746\times10^{-3}$ & $1.578\times10^{-2}$\\
Case 8 & $1/128$ & $1.919\times10^{-3}$ & $1.702\times10^{-3}$ & $8.484\times10^{-3}$\\
Case 9 & $1/256$ & $6.111\times10^{-4}$ & $4.870\times10^{-4}$ & $7.206\times10^{-3}$\\
\hline
\end{tabular}
\end{center}
\caption{Shape errors for drop-in-a-vortex cases.}
\label{tab:shape error:vortex-in-a-box}
\end{table}
Shape errors with three different choices of characteristic lengths are computed for four different grid resolution. Simulation results are provided in Table~\ref{tab:shape error:vortex-in-a-box}. The definitions of shape errors follow the previous analysis except that $\mathcal{E}(\Delta_{32})$ is the shape error with a constant $\Delta_f = 1/32$. In other words, we are fixing the thickness of the interface by selecting $\Delta_f = 1/32$ for four different grid resolutions.

\begin{figure}
    \centering
    \includegraphics[width=\textwidth]{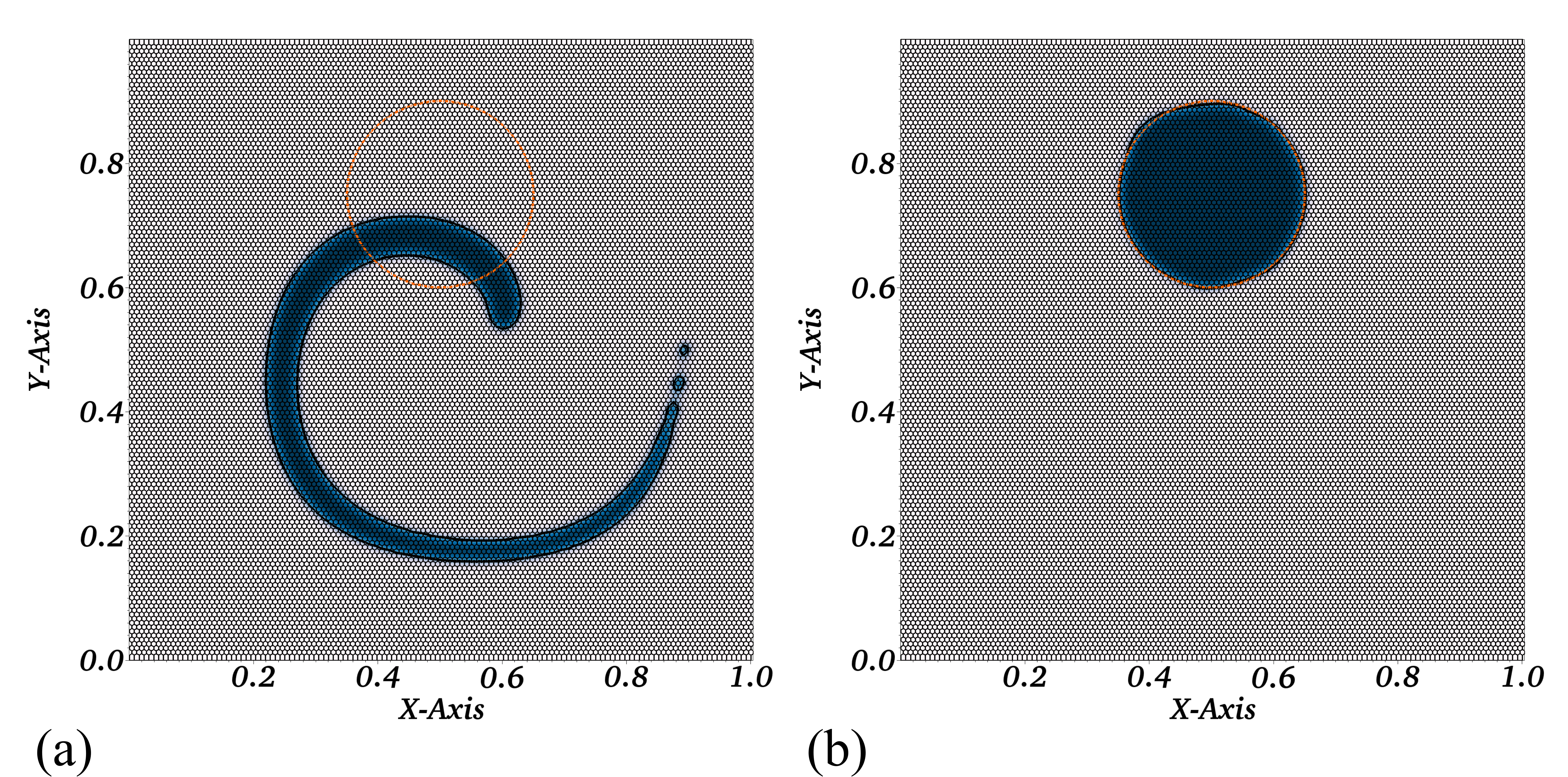}
    \caption{Snapshots of $\phi$ field for a drop-in-a-vortex test case for case 8 in Table~\ref{tab:shape error:vortex-in-a-box} with $\Delta_f = \Delta_{\text{const}}$. The orange solid line shows the initial drop. The blue and white color field indicate $\phi$ values of unity and zero, respectively, at time (a) $t=2$ and (b) $t=4$. The black solid line represents the mesh.}
    \label{fig:fig2}
\end{figure}

Snapshots with the droplet shapes from one of the simulations (case $8$ with constant characteristic length $\Delta_{\text{const}}$) are shown in Figure~\ref{fig:fig2} at a half-time of $t=2$ and a final time of $t=4$. The drop undergoes deformation until $t=2$, and the flow field is reversed to recover the initial drop shape at $t=4$. Figure~\ref{fig:fig2} shows that the method accurately recovers the final circular shape of the drop on an HCP mesh. For all cases, the total volume is discretely conserved throughout the simulation, and we confirm that the boundedness of $\phi$ is preserved.

\begin{figure}
    \centering
    \includegraphics[width=0.7\textwidth]{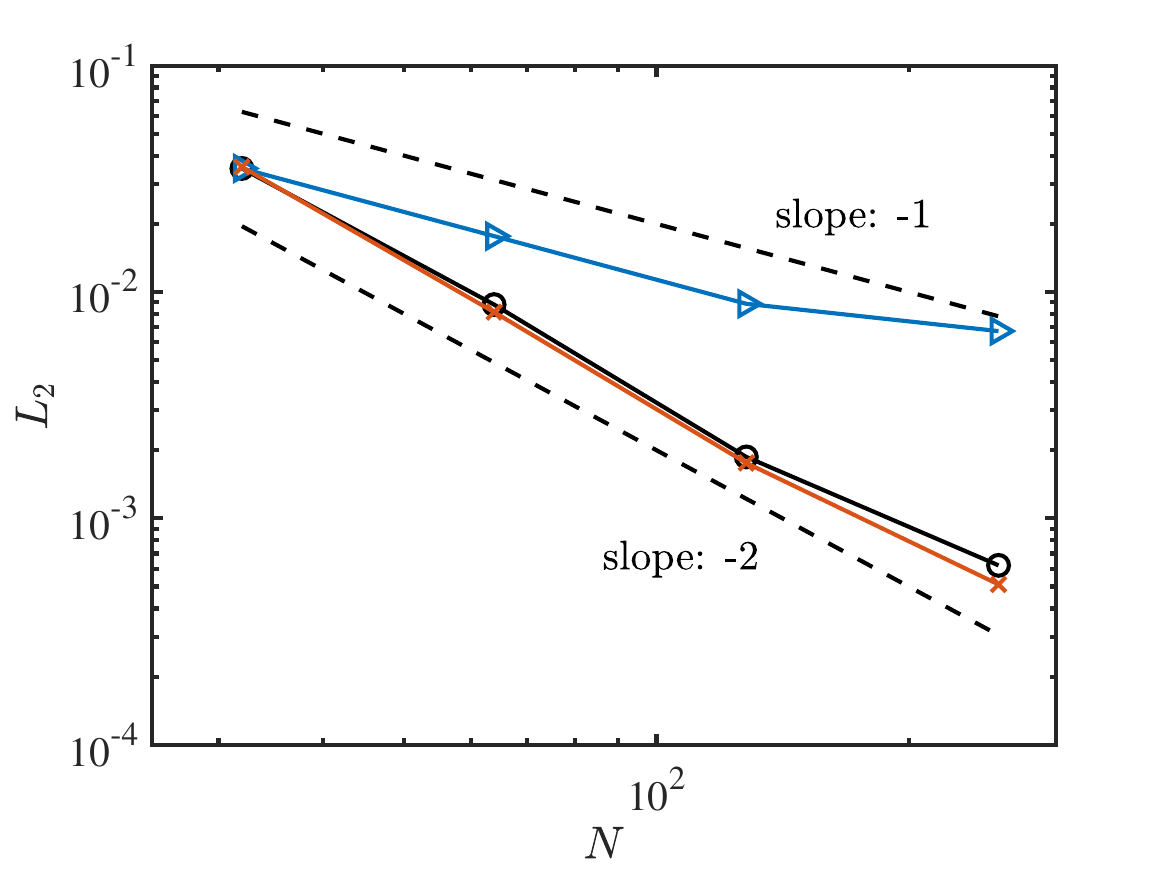}
    \caption{L2-norm error defined in Eq.~\eqref{eq:error metric} for drop in a vortex simulations. Four different uniform HCP grid resolutions are tested with the number of grids in each direction to be $N = 32$, $64$, $128$, and $256$. The non-dimensional ACDI parameters are $\Gamma^* = \Gamma/|\vec{u}|_{max} = 1.0$ and $\epsilon^* = 0.6$. Black-dashed lines show the reference slopes. Lines with symbols refer to simulation results with different choices of $\Delta_f$: $\Delta_f = \Delta_{const}$ (red-cross), $\Delta_f = 1/32$ (blue-triangle), and $\Delta_f = \Delta_{local}$ (black-circles).}
    \label{fig:acdi l2-norm}
\end{figure}

Results in Table~\ref{tab:shape error:vortex-in-a-box} are plotted as a function of the number of grid points $N$ in Figure~\ref{fig:acdi l2-norm} in log scale. The shape errors with constant and local characteristic lengths are almost identical, which is expected because the grid is uniform, and the order of convergence is approximately $2$. For $\Delta_f = 1/32$ case, the order of convergence is approximately $1$. Hence, it is recommended to set $\Delta_f$ to be $\Delta_{const}$ or $\Delta_{local}$. Using $\Delta_f=\Delta_{local}$ might be an easier and a practical option because there is no need to choose a value for $\Delta_f$. In this case, the interface thickness will scale with the local grid size in the domain. 

To compare the performance of the current ACDI approach with the conservative diffuse-interface (CDI) method ~\citep{chiu2011conservative}, the shape errors for the drop-in-a-vortex case using the CDI method is provided in~\ref{sec:app:cdi comparison}. Furthermore, shape errors using Green Gauss gradient reconstruction is provided in~\ref{sec:app:green gauss} as an alternative choice of the gradient reconstruction scheme as opposed to the least squares method in Eq.~\eqref{eq:gradient reconstruction:error minimization 2}.

\subsection{Drop in a homogeneous isotropic turbulence field} \label{sec:result-HIT}
In this section, we perform an infinite and finite Reynolds number simulations of a drop in isotropic turbulence (a test case proposed in \citet{jain2022accurate} to evaluate the nonlinear stability of a method for simulating two-phase turbulent flows). For a stable method, the kinetic energy either remains constant or decays with time. A triply periodic cubic box of size $[0,2\pi]\times[0,2\pi]\times[0,2\pi]$ is considered as the simulation domain. The initial velocity field is set following the model energy spectrum from~\citet{passot1987} as
\begin{equation}
    E(k) \propto k^4 \exp{\left[-2\left(\frac{k}{k_0}\right)^2\right]},
    \label{eq:energy spectrum}
\end{equation}
where $k$ is the wavenumber and $k_0 = 4$ is the most energetic wavenumber. A spherical drop of radius $r = 1.0$ is initially placed at the center of the domain. The dynamic viscosity of the two fluids is chosen to be zero for the infinite Reynolds number case, and the surface tension is set to zero. The simulation is repeated with different density ratios $\rho_1/\rho_2=1$, $10$, $100$, and $1000$.
\begin{figure}
    \centering
    \includegraphics[width=1.0\textwidth]{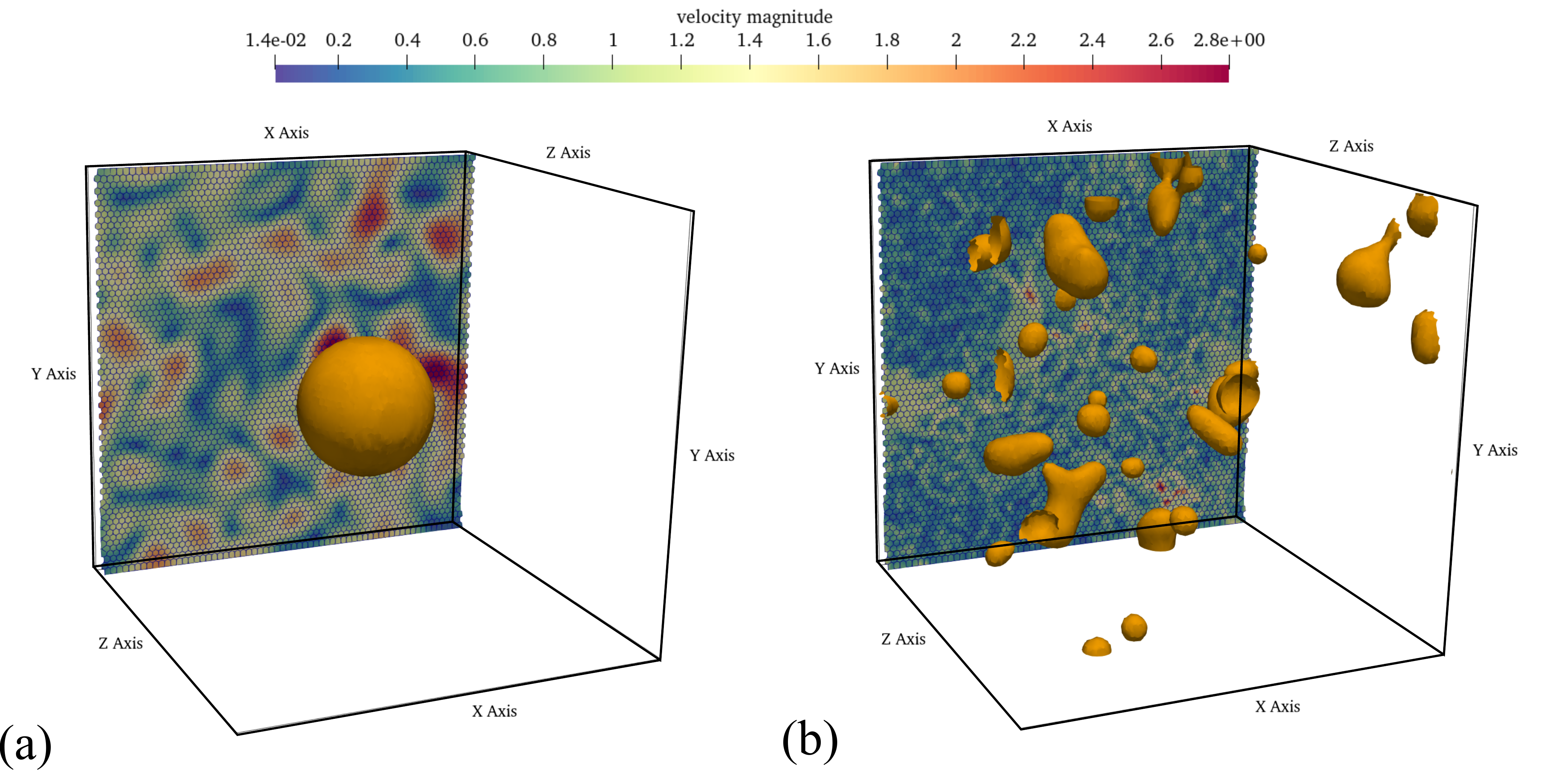}
    \caption{Snapshots of the drop-laden HIT simulations at (a) $t = 0$ and (b) $t = 8$.
    The iso-surface indicates the contour of PF variable with $\phi = 0.5$.}
    \label{fig:fig3}
\end{figure}
The inviscid simulations are run until $t = 8$, and we track the global kinetic energy, which is defined as
\begin{equation}
    K = \sum_{cv} \rho u_{i,cv}u_{i,cv} dV_{cv}.
    \label{eq:total kinetic energy}
\end{equation}
The iso-surface of the PF variable of $\phi = 0.5$ for the density ratio $\rho_1/\rho_2=1000$ is illustrated in Figure~\ref{fig:fig3} at $t=0$ and $t=8$. In all the simulations, the total volume is conserved up to machine precision, and the PF variable is bounded.

\begin{figure}
    \centering
    \includegraphics[width=0.7\textwidth]{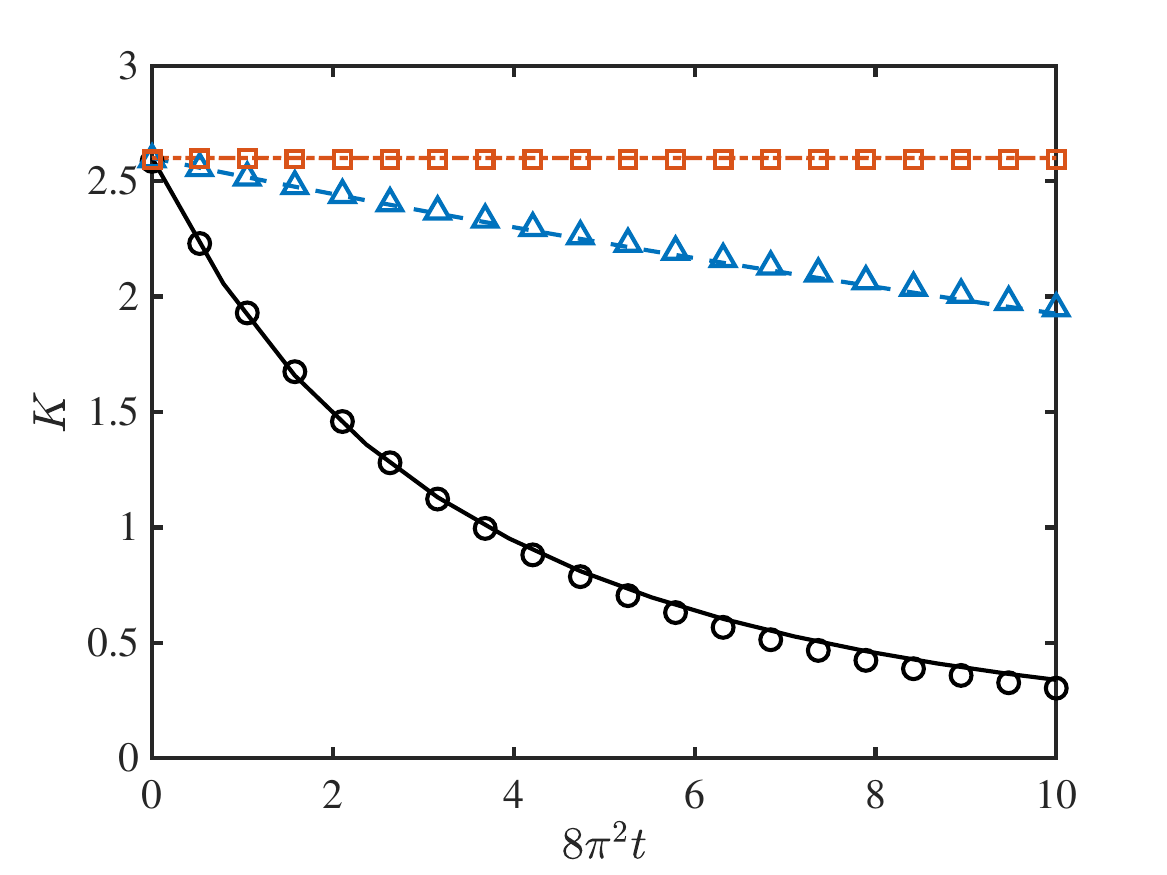}
    \caption{Total kinetic energy decay for isotropic turbulence plotted as a function of time at various Reynolds numbers. Symbols indicate the results obtained by~\citep{mahesh2004numerical} for single-phase flow solver for the Reynolds number $10^6$ (square), $10^4$ (triangle), and $10^2$ (circles). Lines refer to our simulation results for a drop in HIT velocity field with unity density ratio and $\text{Re}=10^6$ (orange-dotted line), $10^4$ (blue-dashed line), and $10^2$ (black-solid line).}
    \label{fig:hit kinetic energy}
\end{figure}

For the unity density ratio case (which consistently reduces to a single-phase simulation), we compare the evolution of global kinetic energy in the present work to~\citet{mahesh2004numerical} in Figure~\ref{fig:hit kinetic energy}. Initial condition and parameters are set identical to~\citet{mahesh2004numerical}.
Here, the Reynolds number is defined based on the Taylor microscale, $\lambda$, as $Re = u_{\text{rms}}\lambda/\nu$, where $u_{\text{rms}}$ is computed as $((2/3)\int^{\infty}_0 E(k) \textnormal{d}k)^{1/2}$.
For Reynolds numbers $Re = 10^2$, $10^4$, and $10^6$, our simulation results show good agreements.

\begin{figure}
    \centering
    \includegraphics[width=0.7\textwidth]{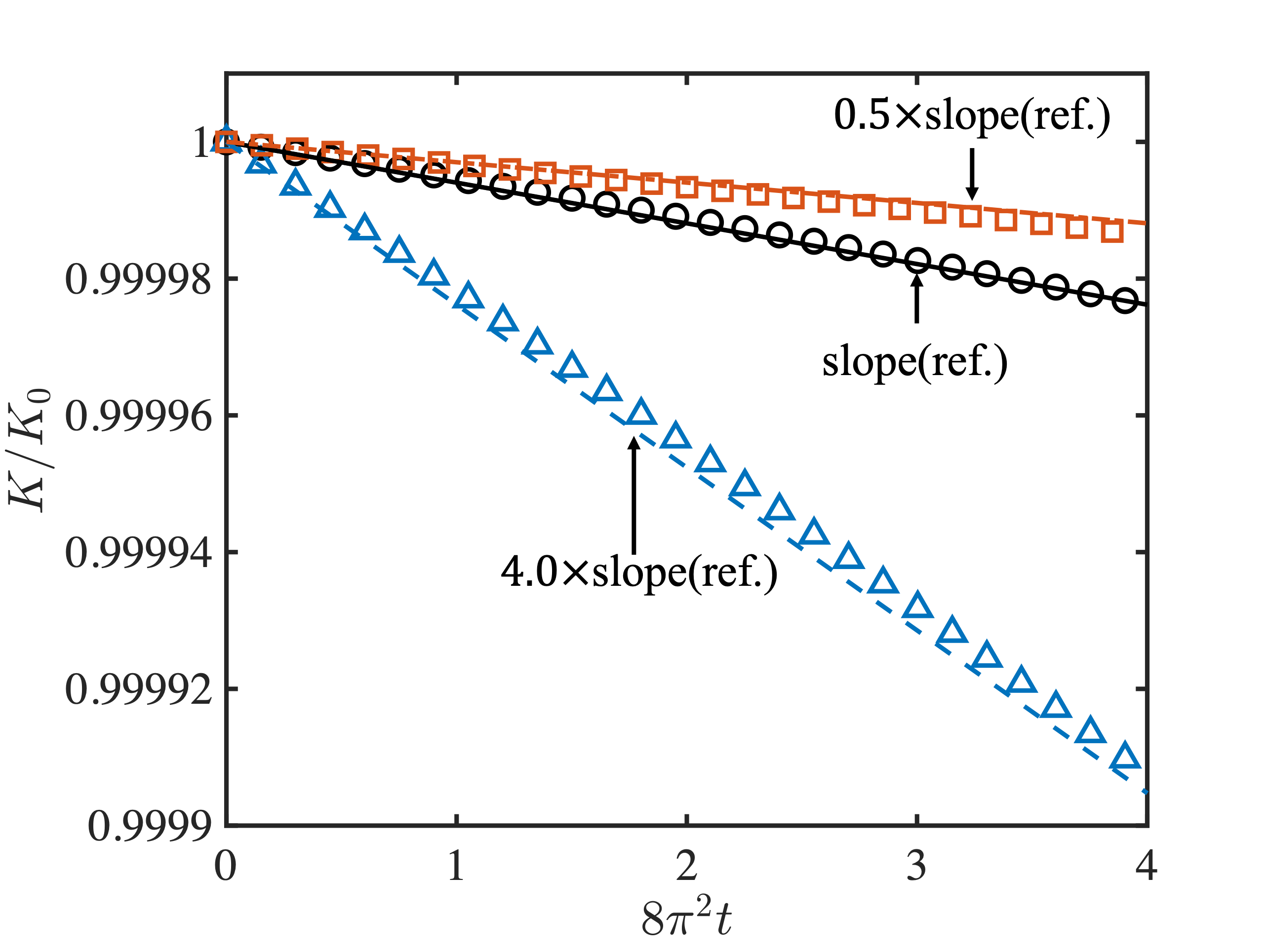}
    \caption{The dissipation of the global kinetic energy in Eq.~\eqref{eq:total kinetic energy} for a drop in HIT velocity field. A reference case with density ratio $\rho_{1}/\rho_{2} = 1000$ is shown using black circles. Two additional cases with the same density ratio, but with half the grid size and half the time step are denoted with blue triangles and orange squares, verifying the scaling factor in Eq.~\eqref{eq:energy budget: error analysis 4}. Solid lines show the approximated slope for each corresponding data.}
    \label{fig:hit kinetic energy error analysis}
\end{figure}

For the infinite Reynolds number simulation, no dissipation mechanisms are present, and hence theoretically the initial global kinetic energy is expected to be preserved. However, we showed in Eq.~\eqref{eq:energy budget: error analysis 4} in Section~\ref{sec:energy-stable collocated grid formulation} that the pressure term in the fractional-step method introduces error in the global kinetic energy transport and this error is strictly dissipative in nature with the novel formulation presented in this work, making the formulation energy stable.
Accordingly, we observe a small decay of kinetic energy in time. This behavior is observed for all the density ratio cases. 
The dissipative term in Eq.~\eqref{eq:energy budget: error analysis 4} drains kinetic energy proportional to the time step size $\Delta t$ and the square of the grid size $\Delta_f^2$ in the limit of early stage of flow evolution, where our assumptions in the derivation of Eq.~\eqref{eq:energy budget: error analysis 4} are still valid. To verify this behavior, we setup a reference case of drop-laden HIT with density ratio $1000$, which is illustrated as a black symbols in Figure~\ref{fig:hit kinetic energy error analysis}. When two additional cases with half the grid size $\Delta_f/2$ (blue solid line) and half the time step $\Delta t/2$ (orange solid line) as the reference case are run, their slope of initial kinetic energy decay follows the scaling of the factor $\Delta t \Delta_f^2$ as derived in Eq.~\eqref{eq:energy budget: error analysis 4}.

Since the error due to pressure terms in the fractional-step method is strictly dissipative in nature, as indicated by the monotonic decay of the global kinetic energy in Figure~\ref{fig:hit kinetic energy error analysis}, this verifies that the proposed method is energy stable. 
Hence, the proposed method is provably robust for all density ratios at infinite $Re$. At finite Re, the method will be stable because of the physical viscosity, which further acts to stabilize the approach.

In~\ref{sec:app:green gauss}, we present the global kinetic energy evolution when the least squares method in Eq.~\eqref{eq:gradient reconstruction:error minimization 2} is replaced by the Green Gauss gradient reconstruction scheme.
In addition, sensitivity of the global kinetic energy dissipation to the constant coefficient $\alpha$ in Section~\ref{sec:projection method algorithm} and the density ratio are presented in~\ref{sec:app:pressure fraction} and~\ref{sec:app:density ratio}, respectively.

\section{Complex two-phase flow applications\label{sec:complex two phase flow apps}}
This section presents three simulations, using the method proposed in this work, that are further used as verification and validation cases for complex two-phase flow applications: a damped surface wave~\citep{Prosperetti1981} in Section \ref{sec:surface-wave}, a liquid jet with optimized disturbance~\citep{hwang2021} in Section \ref{sec:jet-distortion}, and a liquid jet atomization from a non-cavitating ECN's Spray A nozzle~\citep{desantes2016coupled} in Section \ref{sec:spray-A}.

\subsection{Surface damped wave \label{sec:surface-wave}}

\begin{figure}
    \centering
    \subfloat[]{\includegraphics[width=0.35\textwidth]{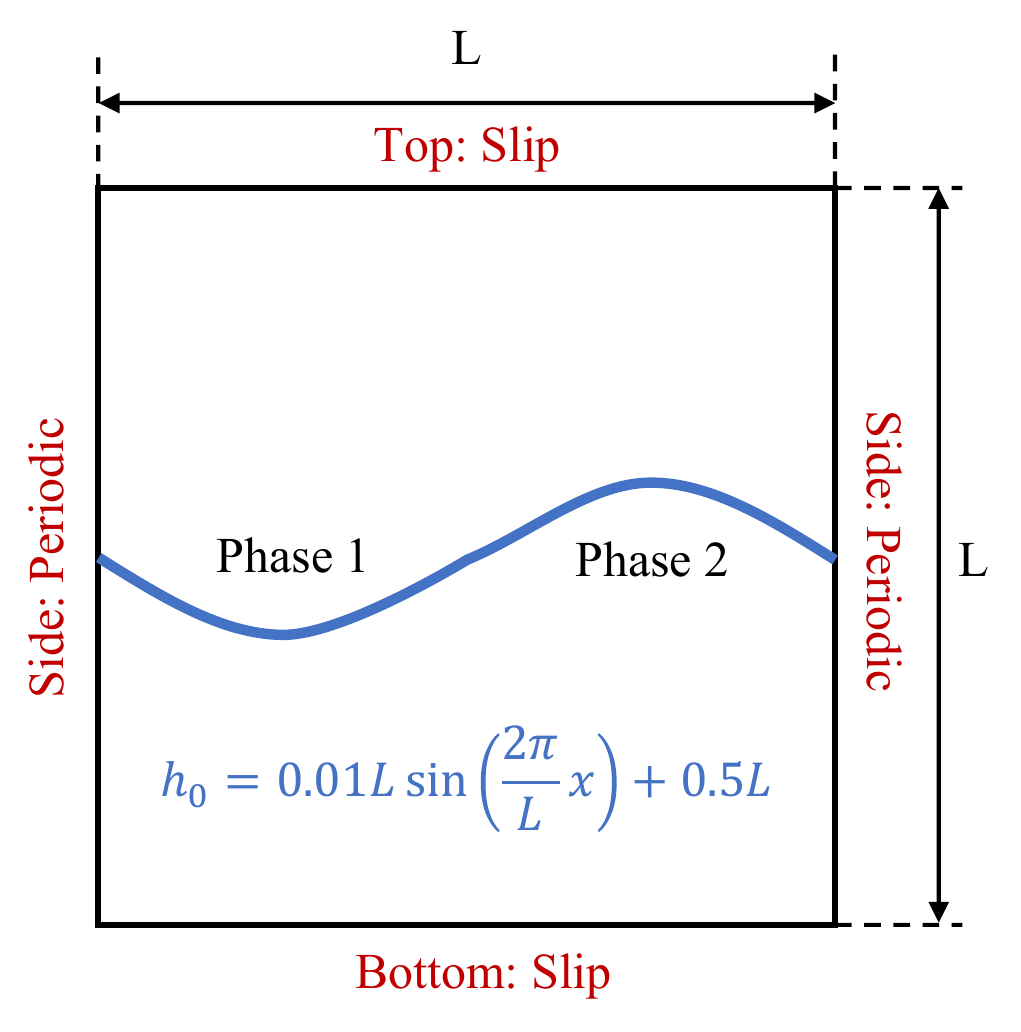}}
    \subfloat[]{\includegraphics[width=0.6\textwidth]{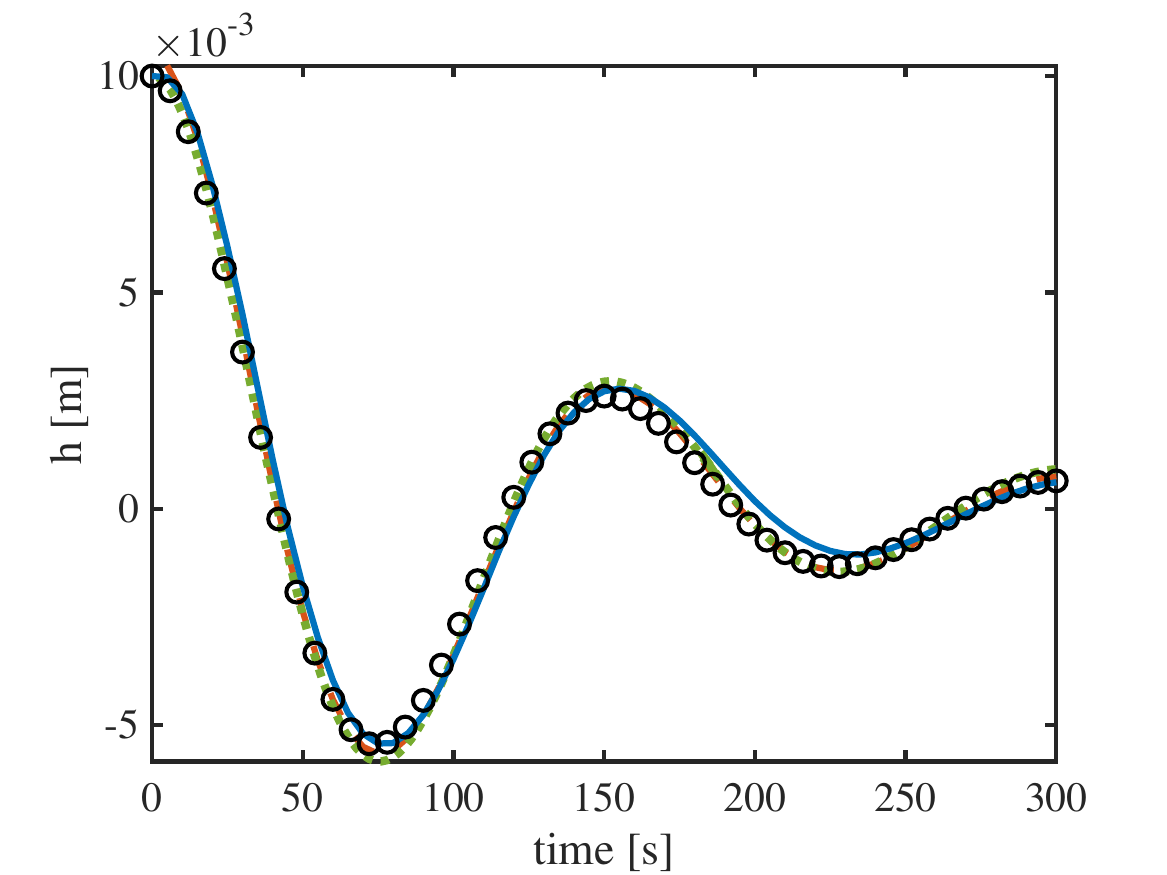}}

        \caption{The damped surface wave simulation. (a) A schematic of the initial problem setup. (b) Wave height as a function of time. Circles denote analytical solution from~\citep{Prosperetti1981}. Lines refer to our simulation results with $\Delta x = L/32$ (blue-solid line), $L/48$ (red-dashed line), and $L/64$ (green-dotted line).}
    \label{fig:damped surface wave}
\end{figure}

In this section, a two-dimensional simulation of a damped surface wave in a domain of size $L \times L$ is presented. The schematic of the domain is shown in Figure~\ref{fig:damped surface wave}(a). Flat interface is perturbed with a sinusoidal wave with $h(x,y) = A \sin(2\pi x/L) + L/2$ with its amplitude $A = 0.01 \times L$.
The phase indicator $\phi$ field is initialized as $\phi = 1/2(1.0 + \tanh{(d/(2*V_{cv}^{1/3}))})$, where $d$ is the minimum distance to interface.
Periodic boundary conditions are prescribed at the side faces and slip boundary conditions are set at the top and bottom faces. Three different grid resolutions, $\Delta x = L/32$, $L/48$, and, $L/64$, with uniform HCP mesh were used.
The density of the two phases are: $\rho_{\text{phase1}} = 1\ kg/m^3$ and $\rho_{\text{phase2}} = 1000\ kg/m^3$. Likewise, viscosity ratio is set to be $1000$, with $\mu_{\text{phase1}} = 6.4720863\times10^{-3}\ Pa\cdot s$ and $\mu_{\text{phase2}} = 6.4720863\ Pa\cdot s$. Surface tension is chosen to be $\sigma = 2.0\ N/m$.
The results are shown in Figure~\ref{fig:damped surface wave}(b), where circles indicate the analytical solution based on initial value theory~\citep{Prosperetti1981}, and lines indicate our simulation results with various grid sizes.

We note that the initial amplitude of the wave is not fully resolved for all choices of the grid resolutions. For example, the initial wave height is only $64\%$ of a grid size, even for the finest case ($\Delta x = L/64$).
Albeit this, simulation results for all grid resolutions show excellent agreements to the analytical solution, except that the lowest resolution $\Delta x = L/32$ simulation demonstrates a minor discrepancy with the analytical solution after $t \approx 150$.

\subsection{Infinitesimal interface distortion on liquid jet \label{sec:jet-distortion}}

As a next validation case, we investigate the effect of an infinitesimal disturbance at the inlet of a liquid jet. The prescribed perturbation within the liquid jet is amplified, and this evolution is attributed to the transient growth mechanism, which eventually leads to interface distortion of the liquid jet. Initial evolution of the liquid jet disturbances is predicted by the linear stability theory~\citep{hwang2021}, providing a reference to the simulations results. Therefore, the current test case demonstrates the ability of the proposed method to capture both the interaction of the disturbance modes and its manifestation on the liquid jet surface.

\begin{figure}
    \centering
    \includegraphics[width=\textwidth]{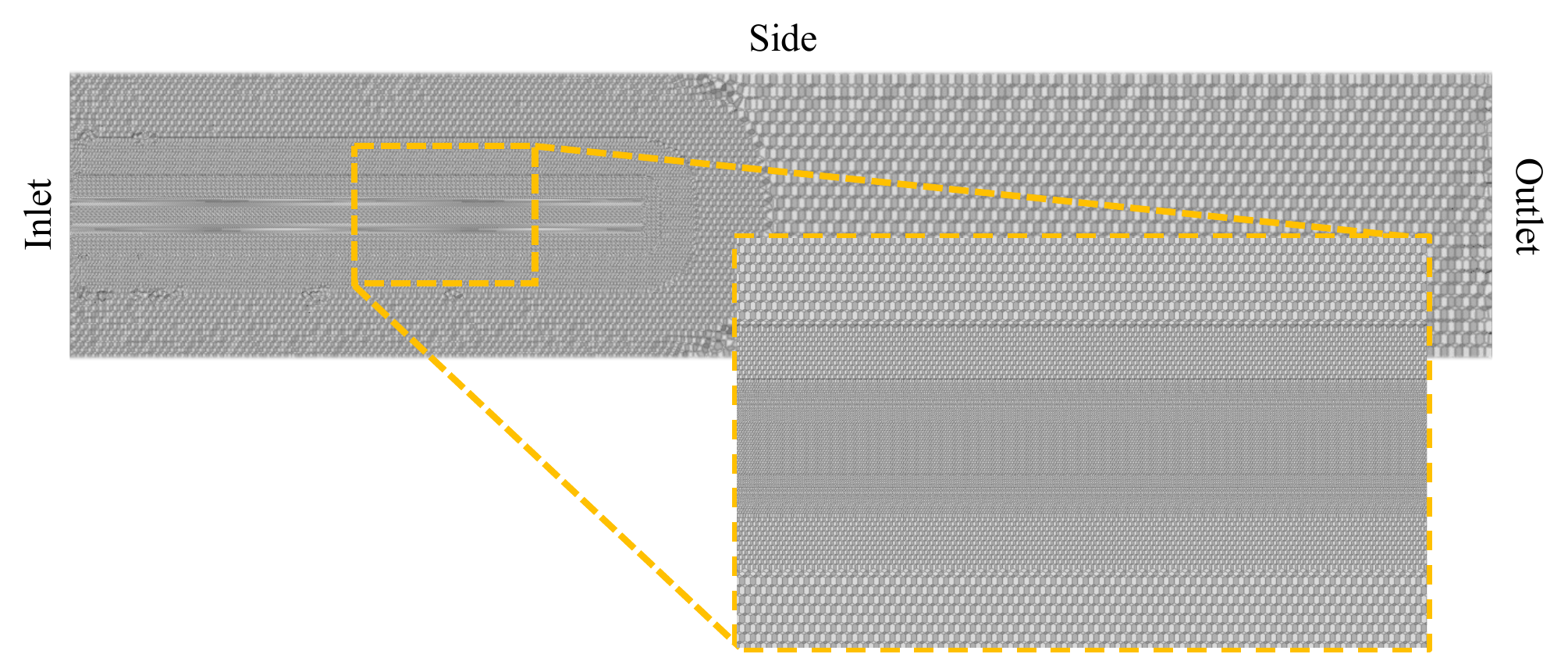}
    \caption{A cross-section of the finer mesh (Mesh $2$) used for the simulation of a liquid jet with infinitesimal interface distortion.}
    \label{fig:thesis mesh}
\end{figure}

A liquid jet is introduced in a cylindrical domain of radius $R = 10R_0$ and length $L = 100R_0$, where $R_0$ is the radius of the liquid jet. The background HCP mesh has a grid size of $\Delta x_{\text{background}} = 1.0R_0$ and is further refined locally with two and three different refinement levels as shown in Figure \ref{fig:thesis mesh} to generate two meshes (Mesh $1$ and Mesh $2$). In Mesh 1, the background coarse grid locally refines into an annular region with its inner radius $R_i = 0.6R_0$, outer radius $R_o = 1.2R_0$, and length $L_c = 40R_0$ from the inlet. The refined zone has a grid size of $\Delta x_{\text{min}} = \Delta x_{\text{background}} / 2^{5}$. The grid transition from the refined region to the background coarse mesh happens over $5$ discrete layers of $20$ cells per each transition layer, where the cells are subsequently coarsened by a factor of $2$ in each layer. 
In Mesh $2$, there is an additional annular zone of refinement with inner radius $R_{ii} = 0.9R_0$, the outer radius $R_{oo} = 1.1R_0$, and the length $L_{cc} = 40R_0$ where the grid size is $\Delta x_{\text{min}} = \Delta x_{\text{background}} / 2^{6}$. 
The total number of cells in Mesh $1$ and Mesh $2$ are $12.6\times 10^6$ and $23.7 \times 10^6$, respectively. A cross-section of the grid for Mesh $2$ is shown in Figure~\ref{fig:thesis mesh}.

Slip boundary condition is prescribed for the side walls and convective outlet boundary condition is chosen for the outlet boundary. At the inlet, the liquid jet velocity and the initial disturbance computed using the linear theory~\citep{hwang2021scale} is prescribed and is shown in Figure~\ref{fig:thesis verification, initial disturbance}. Initial amplitude for the interface disturbance corresponding to disturbance velocities in Figure~\ref{fig:thesis verification, initial disturbance} is $f_0/R_0 = 1.1195$. When the initial disturbance is introduced in the simulations, the magnitudes are scaled by a factor of $10^{-3}$ to make sure that the disturbances are infinitesimal. Parameters are chosen such that the Reynolds number $Re = U_c R_0 / \nu = 7784$ and Weber number $We = \rho U_c^2 R_0 / \sigma = 4865$, where $R_0$ and $U_c$ are the the jet radius and centerline velocity, respectively. The viscosity ratio $m = 0.036$, density ratio $\eta=0.0012$, azimuthal wavenumber $n=6$, and disturbance frequency $\omega = 2.5$. 
For more detailed information, readers are referred to~\citet{hwang2021scale}.

\begin{figure}
    \centering
    \includegraphics[width=\textwidth]{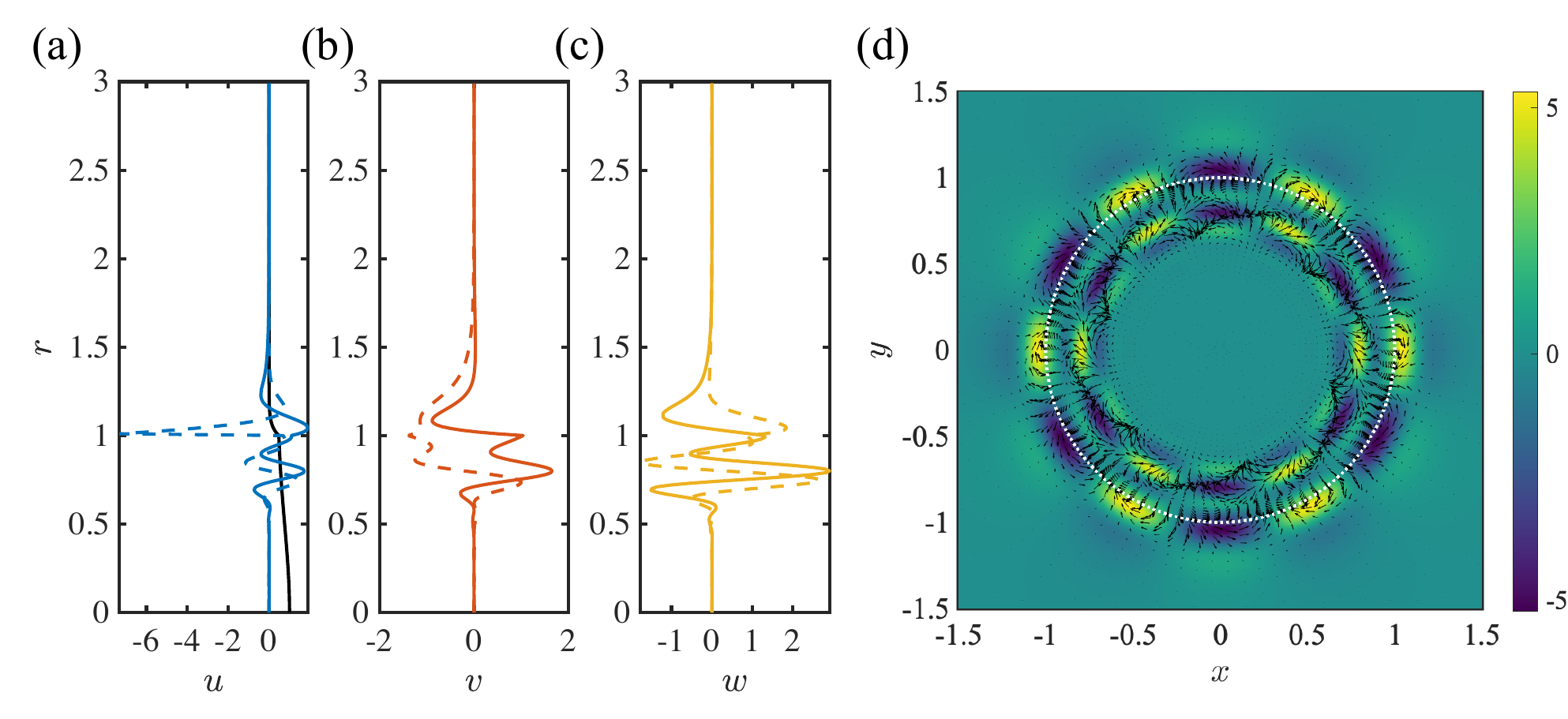}
    \caption{Initial disturbance at the inlet in (a) streamwise, (b) radial, and (c) azimuthal directions. The black solid line in (a) shows the base flow. Solid lines and dashed lines denote the real and imaginary part of the initial disturbance, respectively. For each figure, length is normalized by the jet radius $R_0$ and the velocity is normalized by the centerline velocity $U_c = 1$. (d) A colormap showing the initial disturbance distribution at the cross-plane of liquid jet. The white-dashed line indicates the interface location. The parameters are $Re = 7784$, $We = 4865$, $m = 0.036$, $\eta=0.0012$, $n=6$, and $\omega = 2.5$.}
    \label{fig:thesis verification, initial disturbance}
\end{figure}

\begin{figure}
    \centering
    \includegraphics[width=\textwidth]{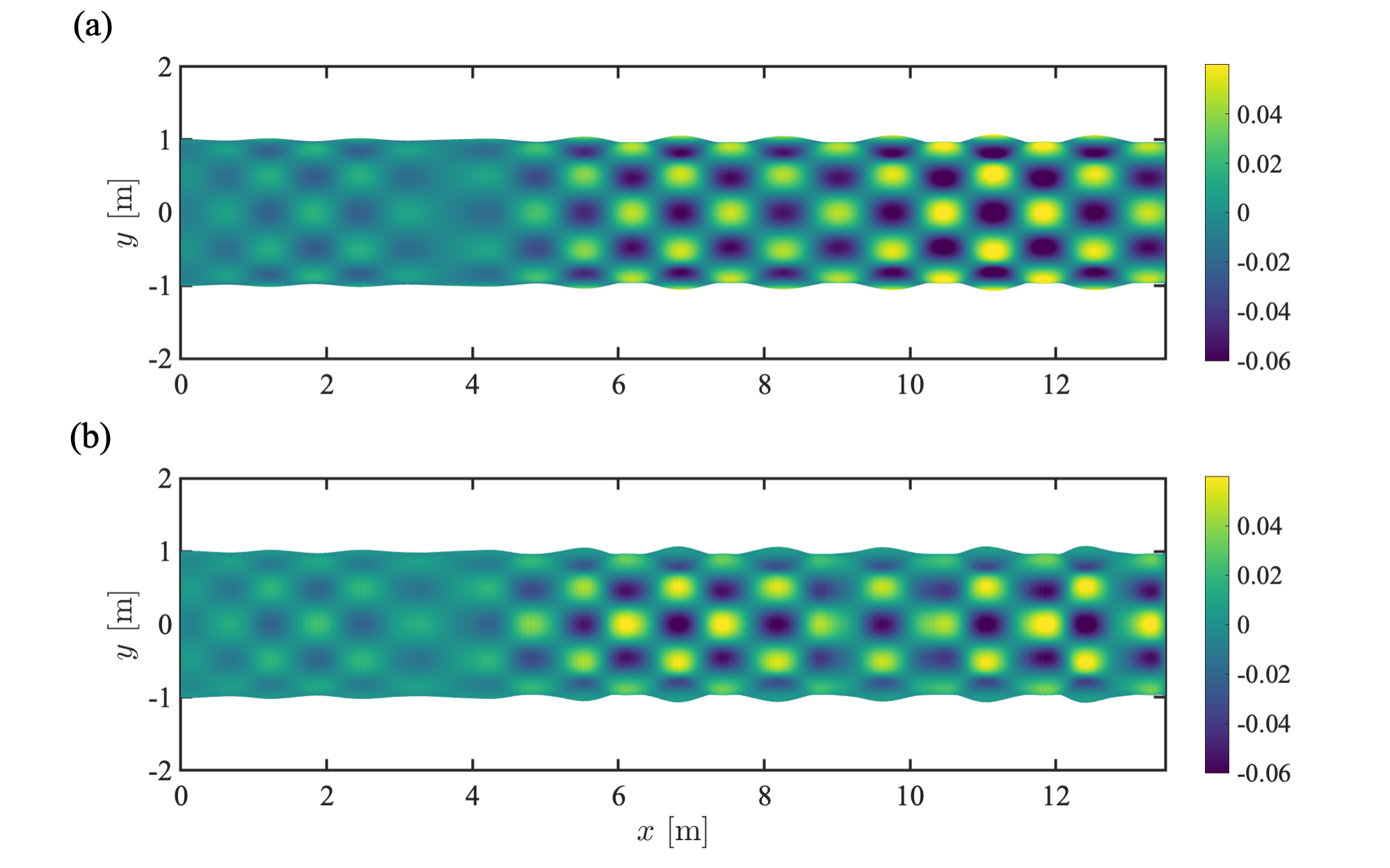}
    \caption{Isosurface of the volume fraction representing the interface of a liquid jet, colored by the local displacement of the interface for the results based on (a) linear stability, and (b) simulations using the proposed method with the finer grid (Mesh $2$). The parameters are $Re = 7784$, $We = 4865$, $m = 0.036$, $\eta=0.0012$, $n=6$, and $\omega = 2.5$.}
    \label{fig:thesis verification, qualitative}
\end{figure}

\begin{figure}
    \centering
    \includegraphics[width=0.7\textwidth]{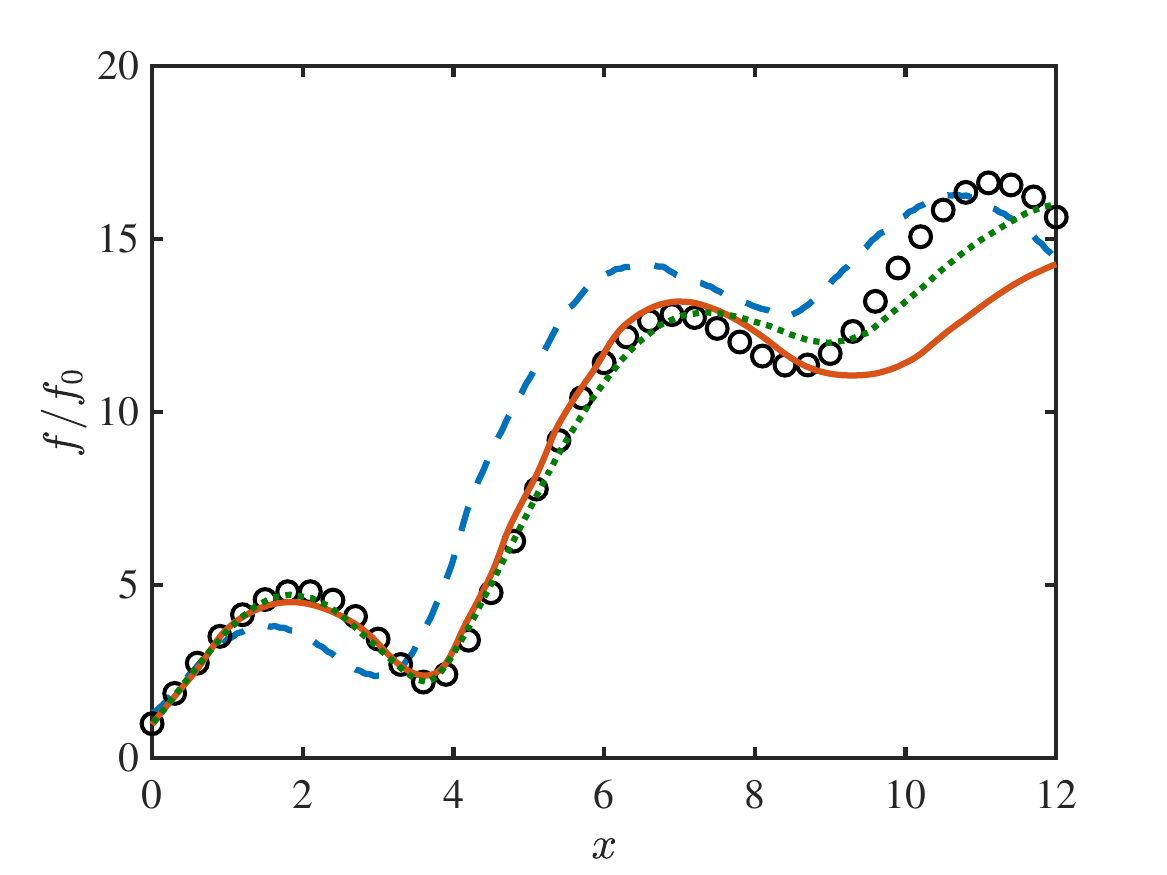}
    \caption{Envelope of the interface distortion of the liquid jet. Circles indicate the reference data based on linear stability theory. Lines refer to low-resolution grid (Mesh $1$) with ACDI method (blue dashed line), high-resolution grid (Mesh $2$) with ACDI method (orange solid line), and open-source geometric VOF solver, Basilisk (green-dotted line). The parameters are $Re = 7784$, $We = 4865$, $m = 0.036$, $\eta=0.0012$, $n=6$, and $\omega = 2.5$.}
    \label{fig:thesis verification, quantitative}
\end{figure}

Both qualitative and quantitative comparison between the linear stability theory prediction and simulation results are provided in Figures~\ref{fig:thesis verification, qualitative}--\ref{fig:thesis verification, quantitative}. As a qualitative comparison, we present the liquid jet interface colored by the local amplitude of the disturbance in Figure~\ref{fig:thesis verification, qualitative}. Again, we note that all the length scales are normalized by the liquid jet radius $R_0$. Simulation prediction in Figure~\ref{fig:thesis verification, qualitative}(b) with the finer grid (Mesh $2$) agrees well with the linear stability theory prediction in Figure~\ref{fig:thesis verification, qualitative}(a) up to $x \approx 8$, beyond which the simulations begin to deviate from the theoretical prediction due to the amplified nonlinear effects.

Figure~\ref{fig:thesis verification, quantitative} shows the quantitative comparison of the envelope of the interface distortion. With the finer grid (Mesh $2$), the prescribed initial perturbation follows linear stability prediction up to $x \approx 8$ as shown with an orange solid line in Figure~\ref{fig:thesis verification, quantitative}. On the other hand, the coarser grid (Mesh $1$) presented as a blue-dashed line deviates from the theoretical prediction much earlier.
In addition, we present the envelope prediction using the open-source code Basilisk, indicated by a green-dotted line~\citep{popinet2015quadtree}. The Basilisk simulation is performed on a uniform Cartesian mesh with cell size $R_0/25.6$, resulting in a total grid count of approximately $10^8$. The Basilisk results exhibit good agreement with the linear theory up to $x \approx 7$ but start to deviate thereafter.

\subsection{Atomization of a liquid jet \label{sec:spray-A}}

In this section, using the ACDI method, we investigate the near-field structure of an atomizing liquid jet  without evaporation. The ECN's Spray A nozzle geometry is utilized as the test case. The Spray A nozzle is known to be asymmetric, and the internal surface of the nozzle possess some level of irregularity. As such, the current application is a good candidate to show the performance of the ACDI method within the Voronoi unstructured-grid framework.

$n$-Dodecane is selected as the liquid fuel, which is injected at $150MPa$ injection pressure into a quiescent environment filled with $N_2$. The injection temperature of the fuel is $343\ K$, the ambient gas temperature is $303\ K$, at which the density of the gas is $22.8\ kg/m^3$.  The inflow of the fuel to the simulation domain through the injector is modeled as a uniform velocity inlet with a velocity of $u_{\text{inlet}} = 13.32\ m/s$, which makes the velocity at the nozzle outlet to be on the order of $O(10^2\ m/s)$. More details about the Spray A geometry and setup can be found in~\citet{desantes2016coupled}.

\begin{figure}
    \centering
    \includegraphics[width=\textwidth]{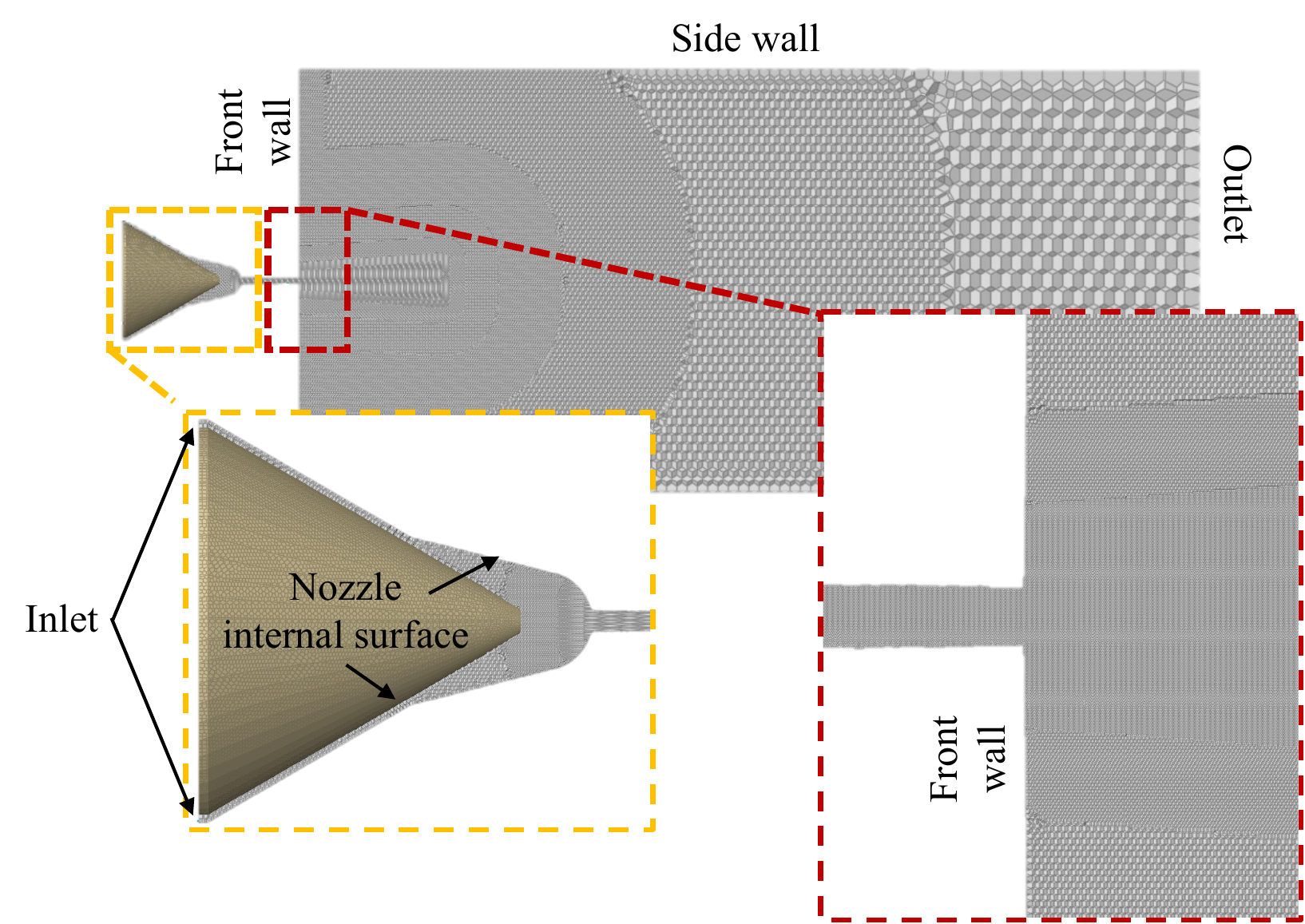}
    \caption{Computational grid for Spray A simulations.}
    \label{fig:sprayA mesh}
\end{figure}

A cylindrical simulation domain is attached at the end of the Spray A nozzle, which is approximately $175\ D_{\text{nozzle}}$ long with $30\ D_{\text{nozzle}}$ in radius, where $D_{\text{nozzle}} = 89.4\times10^{-6}\ m$ is the nozzle outlet diameter. The background HCP grid size is $16\times D_{\text{nozzle}}$. The background mesh is further refined with two local refinement regions. If the center of the nozzle outlet is located at the origin, the first refined region is a box-shaped region ranging
$$\{x,y,z | -10\times10^{-3}\leq x \leq 0.0, -10\times10^{-3}\leq y \leq 10\times10^{-3}, -10\times10^{-3}\leq z \leq 10\times10^{-3}\},$$
in which the grid size is set to $D_{\text{nozzle}}/4$. The second refined region is a T--cone shaped region where the finest grid size is $D_{\text{nozzle}}/32$. The start and end points of the T--cone axis are $(x,y,z) = (-1.05\times10^{-3}, 0, 0)$ and $(2.5\times10^{-3}, 0, 0)$ and the radii at these locations are set to $0.1\times10^{-3}$ and $0.35\times10^{-3}$, respectively. With these refinement regions, the total number of grid points is $46\times 10^6$. A cross-section of the computational grid for the current test case is shown in Figure~\ref{fig:sprayA mesh}.

The computational setup consists of a convective outlet boundary condition at the outlet of the domain, slip on the cylinder at the side wall, an algebraic wall model for the internal surface of the nozzle, and a constant velocity at the nozzle inlet as mentioned earlier. The simulation is run until the total sum of the liquid jet surface area reaches a steady state within the most refined region, e.g., $x \leq 6.0\ mm$.  

\begin{figure}
    \centering
    \includegraphics[width=\textwidth]{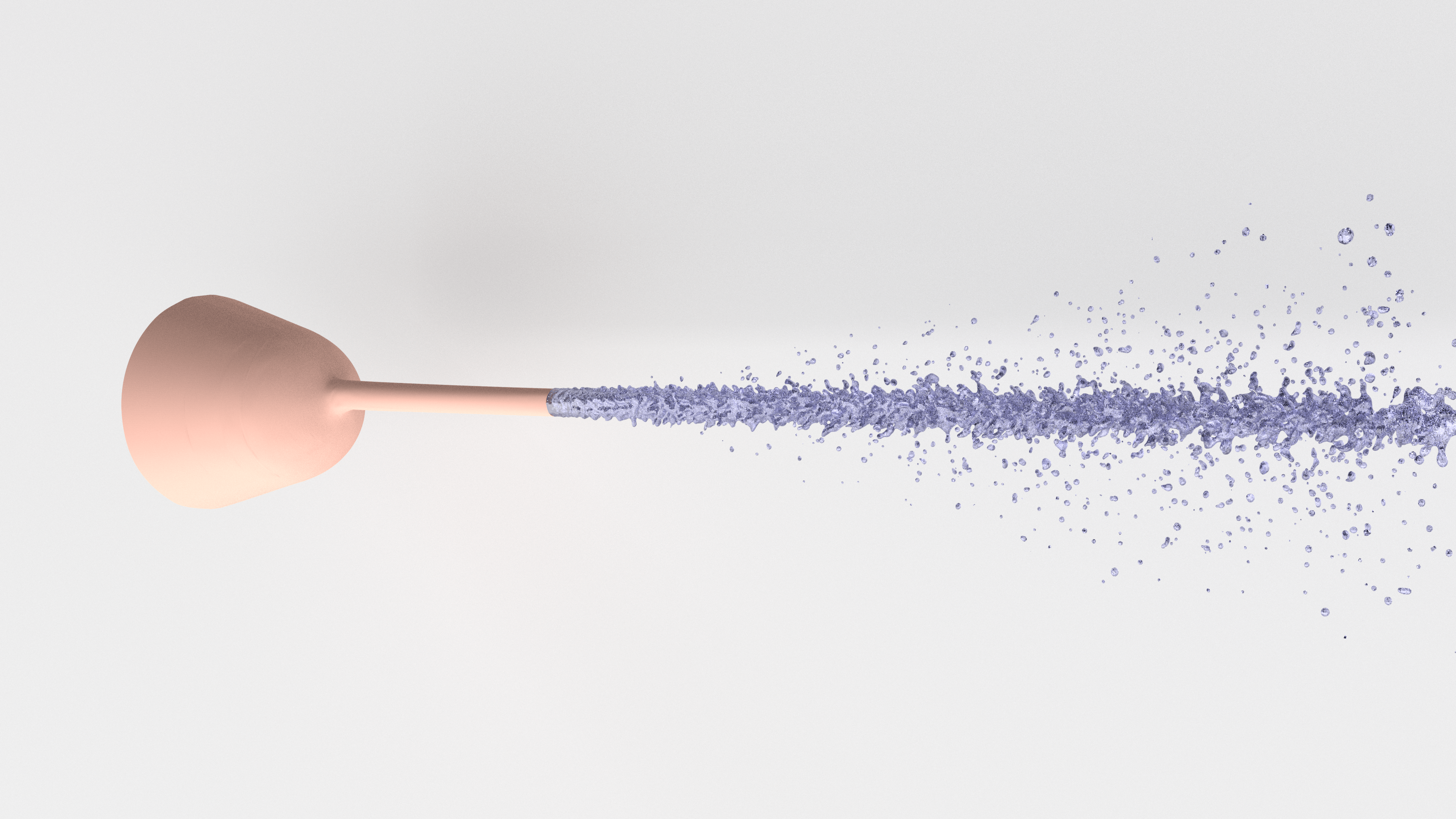}
    \caption{A realistic rendering of a snapshot of Spray A simulation, showing the injector and the liquid jet.}
    \label{fig:sprayA snapshot}
\end{figure}

A snapshot of the Spray A simulation is shown in Figure~\ref{fig:sprayA snapshot} with isosurface ($\phi = 0.5$) of liquid jet. High-speed liquid injection, as well as the irregular inner-nozzle surface and asymmetry, promotes atomization of liquid fuel jet, leading to extensive interface perturbation and droplet generation.

\begin{figure}
    \centering
    \includegraphics[width=0.7\textwidth]{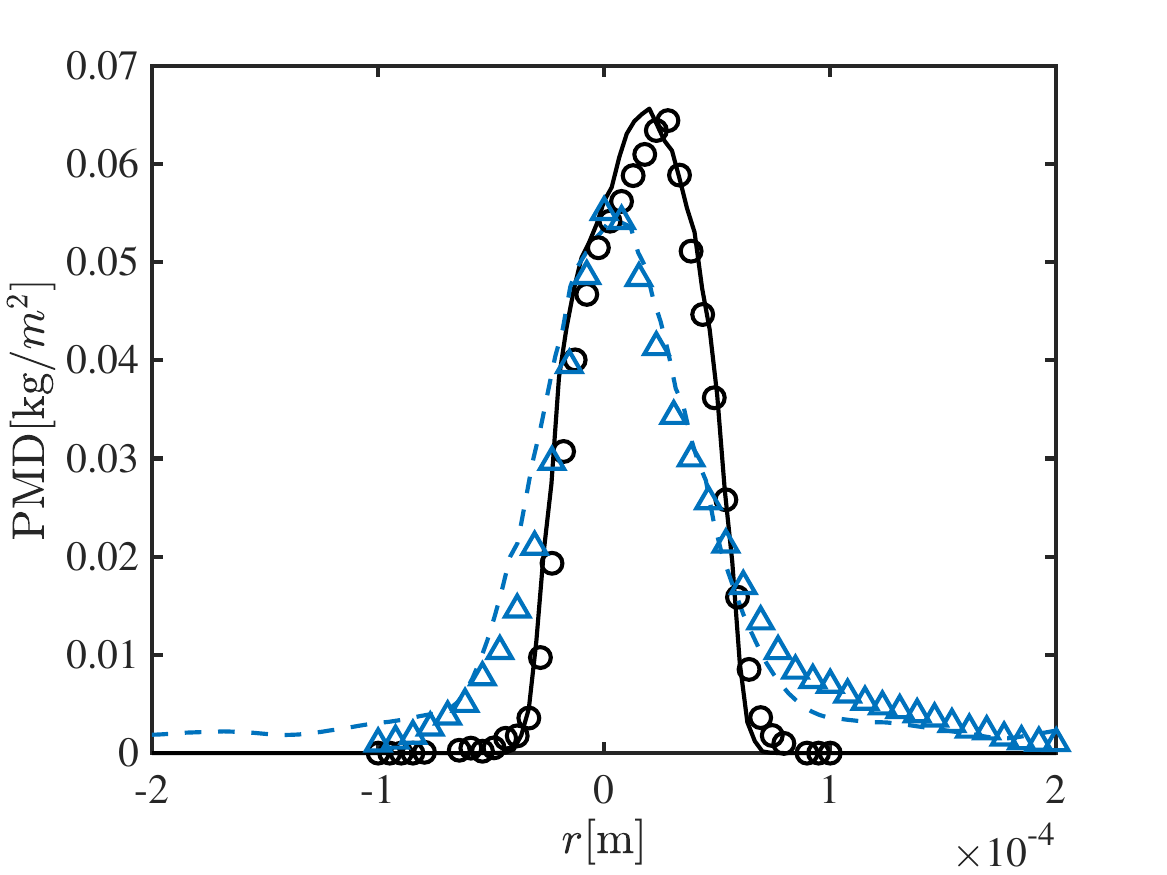}
    \caption{PMD at $250\ \mu s$ at streamwise locations of $0.1\ mm$ (black-solid line) and $6.0\ mm$ (blue-solid line). Circles and triangles indicate the experimental data from~\citep{desantes2016coupled} at $0.1\ mm$ and $6.0\ mm$, respectively.}
    \label{fig:sprayA pmd}
\end{figure}

We measure the projected mass density (PMD) as a quantity of interest at two streamwise locations, $x = 0.1\ mm$ and $x = 6.0\ mm$. The PMD of liquid fuel at these two streamwise locations at $t = 250\ \mu s$ are plotted in Figure~\ref{fig:sprayA pmd}. Note that the measured PMD data depends on the sample plane due to the nozzle asymmetry, and thus, the PMD data on $y$-plane is plotted and compared with the experimental data. This asymmetry is also manifested in the PMD data at both streamwise locations, which are in good agreements with experimental observation.

\section{Conclusions\label{sec:conclusions}}

In this work, an accurate and robust phase-field method is presented for the simulation of two-phase flows on collocated and unstructured grids. We adopt the ACDI method recently proposed by \citet{jain2022accurate} and present a formulation for the Voronoi-mesh-based unstructured grid.
The Voronoi grid is chosen in this study because of its many attractive features, such as uniquely defined connectivity, suitability for the use of non-dissipative central schemes, and highly parallelizable mesh generation.
Because the ACDI method is significantly more accurate than other existing phase-field methods, its application to Voronoi-grid-based unstructured grids has great potential to be an effective interface-capturing method for the simulation of two-phase flows in realistic engineering applications.

We also present a novel consistent fractional-step formulation for collocated grids and show, for the first time, that the formulation is provably energy stable, which will result in a robust formulation for the simulation of two-phase flows. 

To evaluate the applicability of the proposed method for the simulations of two-phase flows on unstructured grids, we first assess the method using the canonical droplet advection and drop in a vortex simulations.
In addition to the use of canonical test cases, we also design new test cases to fully evaluate the performance of interface-capturing methods on unstructured grids with grid transitions. We used these cases to evaluate the accuracy of the proposed method and showed that the proposed method is accurate and is suitable to be used for accurate numerical simulations with unstructured grids on complex domains of engineering interest that will involve grid transitions.
In all the simulations, we confirmed that the volume fraction was bounded without the need for clipping or the use of flux limiter. Moreover, the total volume (mass) was discretely conserved. 

We also performed droplet-laden HIT simulations at infinite Reynolds number to evaluate the robustness of the method for turbulent flow simulations. The proposed method is found to be stable for all the density ratios considered due to the use of consistent momentum equations and the novel consistent fractional-step method proposed in this work. The total kinetic energy was found to decrease slightly with time for all the density ratio cases, verifying that the error from the pressure terms in the proposed fractional-step projection method is strictly dissipative in nature on collocated grids according to the presented proof. Hence, the proposed formulation is provably energy stable for the simulation of two-phase flows.

Finally, we demonstrated a variety of complex simulations for two-phase flow applications including, a surface-damped wave, a spatial evolution of inlet disturbance on a liquid jet interface, and atomization of a liquid fuel in ECN spray A nozzle. These simulations illustrate the suitability of the proposed method for complex simulations of engineering interest, and the advantages of the use of an unstructured framework. These cases showed good agreements to the experimental data or theory on the quantities of interest, thus verifying and validating the approach, while preserving volume conservation and phase-field variable boundedness. The high accuracy, low cost, robustness, and applicability of our approach to complex two-phase applications makes it a promising method for future realistic simulations of two-phase flows.

\section*{Acknowledgments}
This study was supported by funding from the National Science Foundation under grant CMMI-21311961 (to H. H.) and Boeing Co. (to S. S. J.).
The authors would also like to acknowledge the computing time on the Quartz machine at LLNL made available through the Predictive Science Academic Alliance Program (PSAAP) III at Stanford University.

\appendix

\section{Comparison between the ACDI and CDI approaches\label{sec:app:cdi comparison}}
\citet{chiu2011conservative} proposed the conservative diffuse-interface (CDI) model for incompressible two-phase flows by removing the curvature-driven motion from the Allen-Cahn equation. The phase-field variable transport equation for CDI can be written as,
\begin{equation}
\frac{\partial \phi}{\partial t} + \vec{\nabla}\cdot(\phi \vec{u}) = \vec{\nabla}\cdot\left\{\Gamma\left\{\epsilon\vec{\nabla}\phi - \phi (1-\phi) \frac{\vec{\nabla} \phi}{|\vec{\nabla} \phi|}\right\}\right\},
\label{eq:cdi}
\end{equation}
where the variables in Eq.~\eqref{eq:cdi} have identical definitions to the variables in Eq.~\eqref{eq:acdi}. On the RHS of Eq.~\eqref{eq:cdi}, the sharpening term is computed directly in terms of the phase-field variable.
In this section, we use the CDI method on Voronoi unstructured grid and the results are compared with the unstructured ACDI method proposed in this work using two canonical test cases: vortex-in-a-box and a drop in HIT field cases. The constant characteristic lengths are used for grid resolutions from $\Delta = 1/32$, $1/64$, $1/128$, and $1/256$.

\begin{table}
\begin{center}
\setlength{\tabcolsep}{12pt}
\begin{tabular}{c c}
\hline
$\Delta_{\text{const}}$ & CDI\\
\hline
\hline
$1/32$  & $5.375\times10^{-2}$ \\
$1/64$  & $1.755\times10^{-2}$ \\
$1/128$ & $2.743\times10^{-3}$ \\
$1/256$ & $8.304\times10^{-4}$ \\
\hline
\end{tabular}
\end{center}
\caption{Shape errors with constant characteristic lengths $\mathcal{E}(\Delta_{\text{const}})$ for drop-in-a-vortex cases using CDI and least square gradient reconstruction method.}
\label{tab:app:shape error, cdi}
\end{table}

\begin{figure}
    \centering
    \subfloat[]{\includegraphics[width=0.45\textwidth]{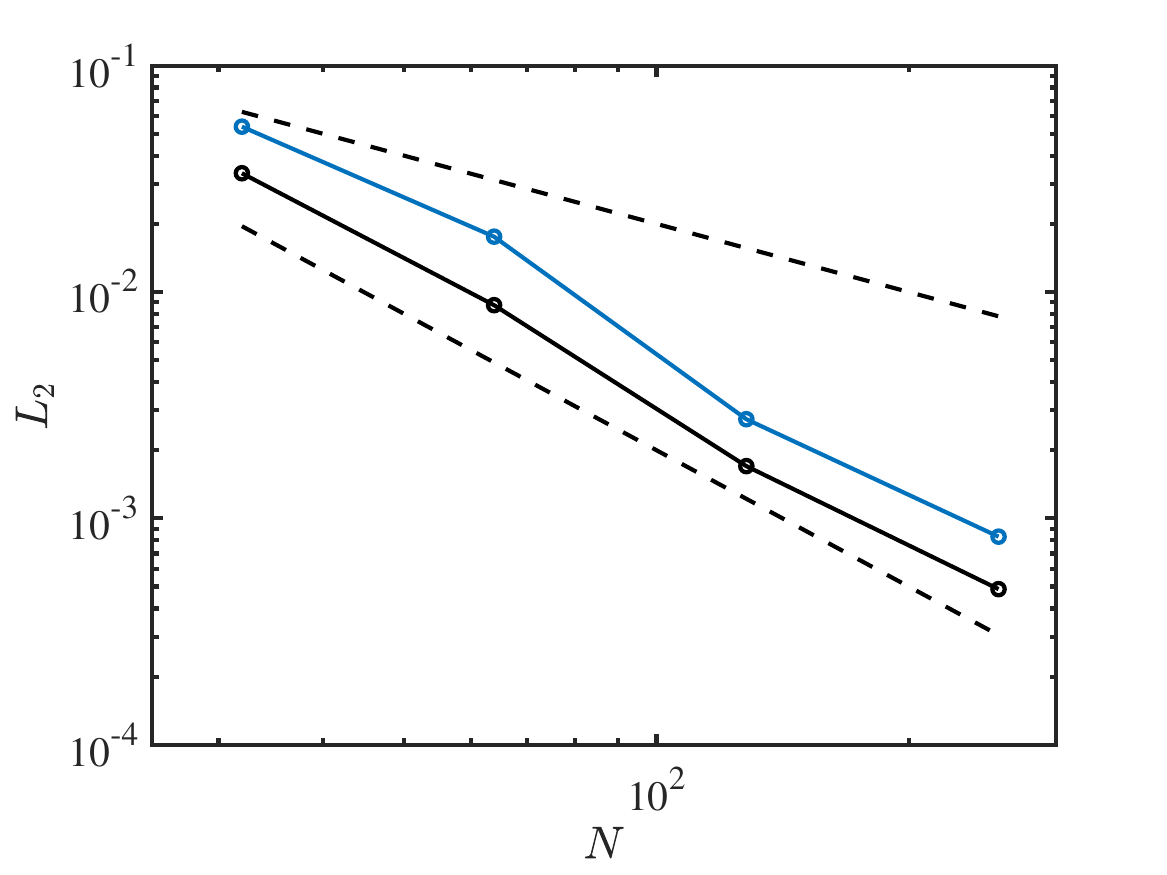}}
    \subfloat[]{\includegraphics[width=0.45\textwidth]{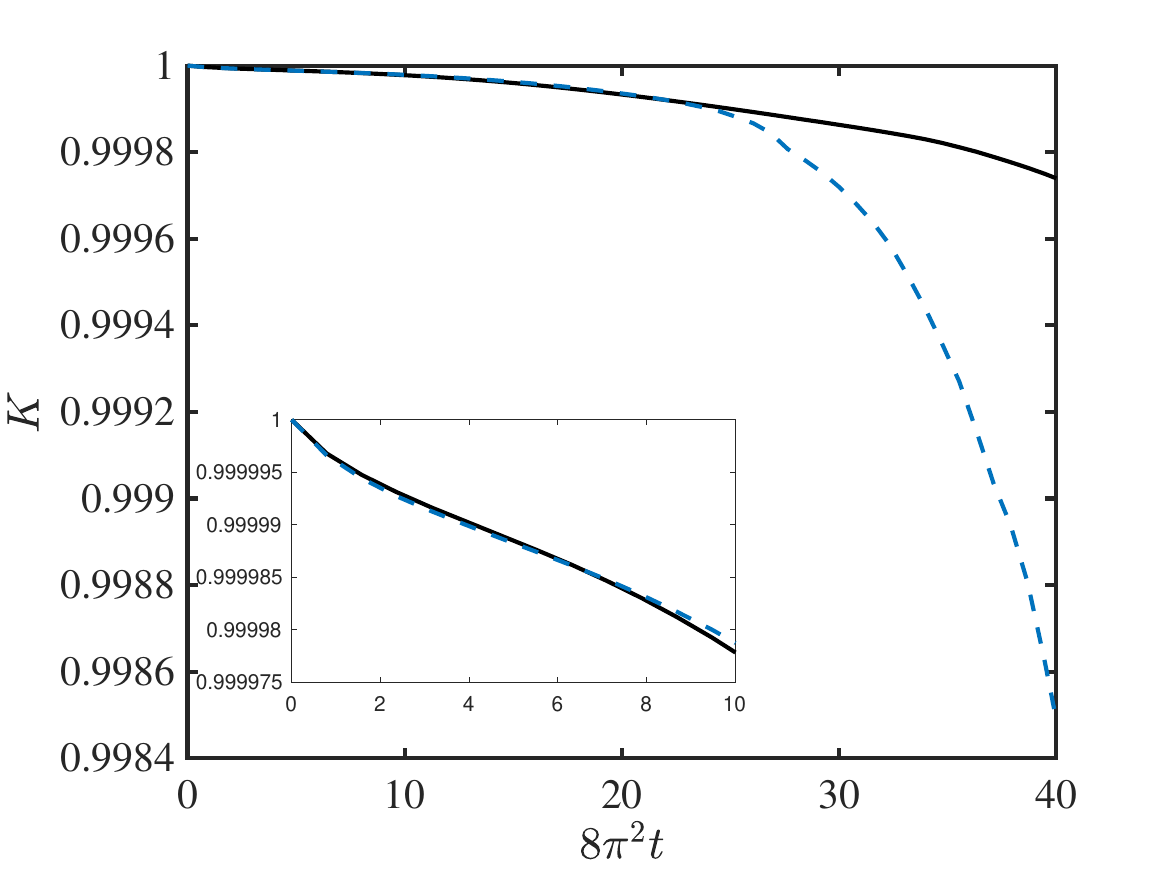}}
    \caption{
    (a) Shape errors from the drop-in-a-vortex cases using CDI method with least square gradient reconstruction are represented by the blue-solid line. The black line represents the reference when current ACDI method and least-square reconstruction are utilized. A density ratio of $\rho_{1}/\rho_{2} = 1000$ and constant characteristic lengths $\Delta_{\text{const}}$ are employed.
    (b) Total kinetic energy dissipation using CDI method with least square gradient reconstruction is depicted by the blue-dashed line. The black line represent the reference when current ACDI method and least-square reconstruction are used, employing a time step of $\Delta t = 0.0002$. A density ratio of $\rho_{1}/\rho_{2} = 1000$ and constant characteristic lengths $\Delta_{\text{const}}$ are utilized.
    }
    \label{fig:app: cdi results}
\end{figure}

Shape errors using the CDI method are presented in Table~\ref{tab:app:shape error, cdi}.
Here, the dimensionless parameters in the interface-regularization term, RHS of Eq.~\ref{eq:acdi}, were chosen as $(\epsilon^* = 0.6$, $\Gamma^* = 5.0)$ that are required to maintain boundedness of $\phi$ with the CDI model. See discussion on boundedness for the CDI model in \citet{Mirjalili2020} (for incompressible flows) and \citet{jain2020conservative} (for compressible flows).
The shape errors obtained using the CDI method exhibit nearly twice the magnitude of errors compared to those of the ACDI method in the present work (see, Table \ref{tab:shape error:vortex-in-a-box}).
Shape errors in Table~\ref{tab:app:shape error, cdi} are depicted in Figure~\ref{fig:app: cdi results}(a) to analyze the convergence rate of the CDI method. As expected, the convergence order for the CDI method is also second order.
The total volume of the droplet is perfectly conserved for both methods, as expected. However, it's worth noting that a smaller time step is required for the CDI method ($\Delta t = 0.0005$) compared to that of the ACDI method ($\Delta t = 0.001$) to maintain boundedness and robustness due to the CFL restriction. %

Furthermore, total kinetic energy decay is investigated using the drop in HIT field case for CDI. The timesteps used for both approaches are $\Delta t = 0.0002$. The result is presented in Figure~\ref{fig:app: cdi results}(b). The CDI method shows comparable kinetic energy decay as the current ACDI approach during the early times of the simulation, however, it demonstrates a sudden abrupt decay at later times making it more dissipative.

\section{Green Gauss gradient reconstruction method\label{sec:app:green gauss}}

\begin{table}
\begin{center}
\setlength{\tabcolsep}{12pt}
\begin{tabular}{c c}
\hline
$\Delta_{\text{const}}$ & Green Gauss\\
\hline
\hline
$1/32$  & $2.976\times10^{-2}$\\
$1/64$  & $7.567\times10^{-2}$\\
$1/128$ & $2.240\times10^{-3}$\\
$1/256$ & $7.265\times10^{-4}$\\
\hline
\end{tabular}
\end{center}
\caption{Shape errors with constant characteristic lengths $\mathcal{E}(\Delta_{\text{const}})$ for drop-in-a-vortex cases using the current ACDI method and Green Gauss gradient reconstruction method.}
\label{tab:app:shape error, green gauss}
\end{table}

To minimize the dissipation error as described in Eq.~\eqref{eq:gradient reconstruction:error minimization 2}, the least square gradient reconstruction, $\lsOp{\cdot}$, has been employed in Section~\ref{sec:projection method algorithm}. This procedure maps the values at cell faces to the values at the cell center.
In this section, we replace this gradient reconstruction method from the least squares approach to the Green-Gauss approach, defined as
\begin{equation}
    \vec{\nabla} \theta(p) \approx \frac{1}{V_{cv}}\sum_{f_k} \theta_{f_k} A_{f_k} \vec{n}_{f_k},
    \label{eq:app:green gausee reconstruction}
\end{equation}
where $\theta$ is an arbitrary variable. 

\begin{figure}
    \centering
    \subfloat[]{\includegraphics[width=0.45\textwidth]{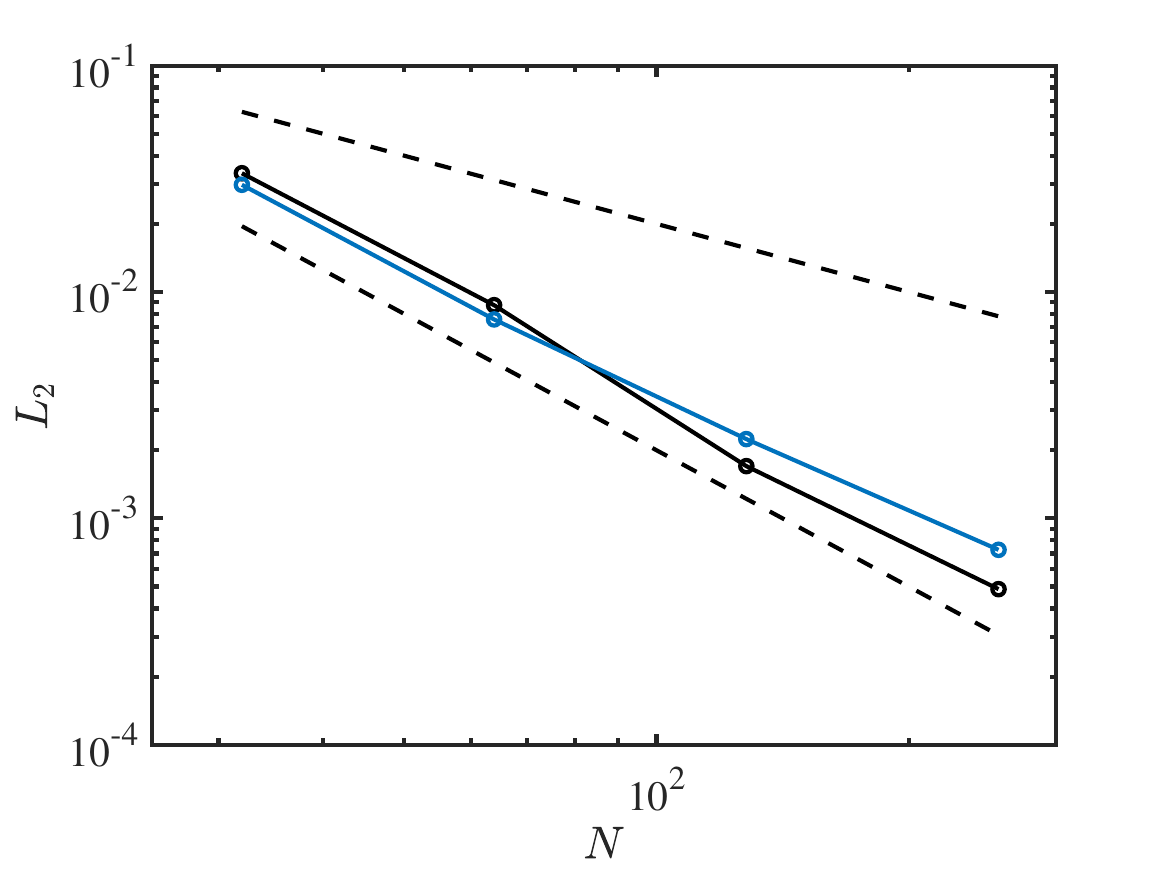}}
    \subfloat[]{\includegraphics[width=0.45\textwidth]{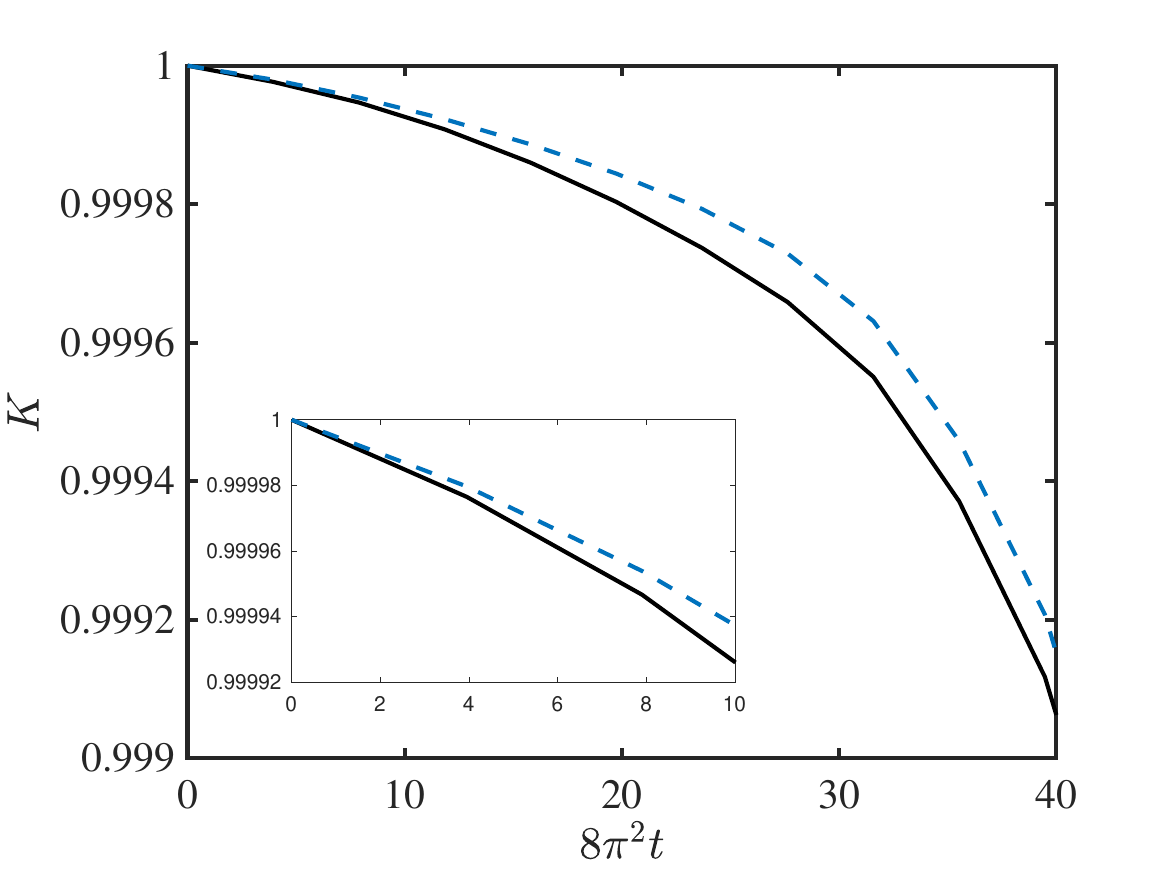}}
    \caption{
    (a) Shape errors from the drop-in-a-vortex cases using the current ACDI method with Green Gauss gradient reconstruction are represented by the blue-solid line. The black line represents the reference when current ACDI method and least-square reconstruction are utilized. A density ratio of $\rho_{1}/\rho_{2} = 1000$ and constant characteristic lengths $\Delta_{\text{const}}$ are employed.
    (b) Total kinetic energy dissipation using the current ACDI method with least square gradient reconstruction is depicted by the blue-dashed line. The black line represent the reference when current ACDI method and least-square reconstruction are used, employing a time step of $\Delta t = 0.001$. A density ratio of $\rho_{1}/\rho_{2} = 1000$ and constant characteristic lengths $\Delta_{\text{const}}$ are utilized.
    }
    \label{fig:app: green gauss results}
\end{figure}

Shape errors using Green Gauss reconstruction in Eq.~\eqref{eq:app:green gausee reconstruction} are reported in Table~\ref{tab:app:shape error, green gauss}.
The timestep used in this test is $\Delta t = 0.001$.
The results using Green-Gauss reconstruction exhibit a comparable magnitude of shape error to the least squares reconstruction (Table \ref{tab:shape error:vortex-in-a-box}), with the maximum difference in shape errors between the two methods being approximately $0.57\%$ and the least squares method showing better convergence compared to the Green Gauss method.
The total volume of the droplet is perfectly conserved, and the phase-field variable $\phi$ is bounded. 
Approximately, a second-order convergence rate is demonstrated in Figure~\ref{fig:app: green gauss results}(a).
Additionally, we investigate the evolution of total kinetic energy using Green-Gauss reconstruction for a drop in a HIT field case. The timestep used in this test is identical to the previous case, $\Delta t = 0.001$. The results are illustrated in Figure~\ref{fig:app: green gauss results}(b). The Green-Gauss reconstruction shows slightly less kinetic energy decay compared to the least squares reconstruction.

\section{Effect of the coefficient $\alpha$ on the global kinetic energy dissipation\label{sec:app:pressure fraction}}
\begin{figure}
    \centering
    \includegraphics[width=0.6\textwidth]{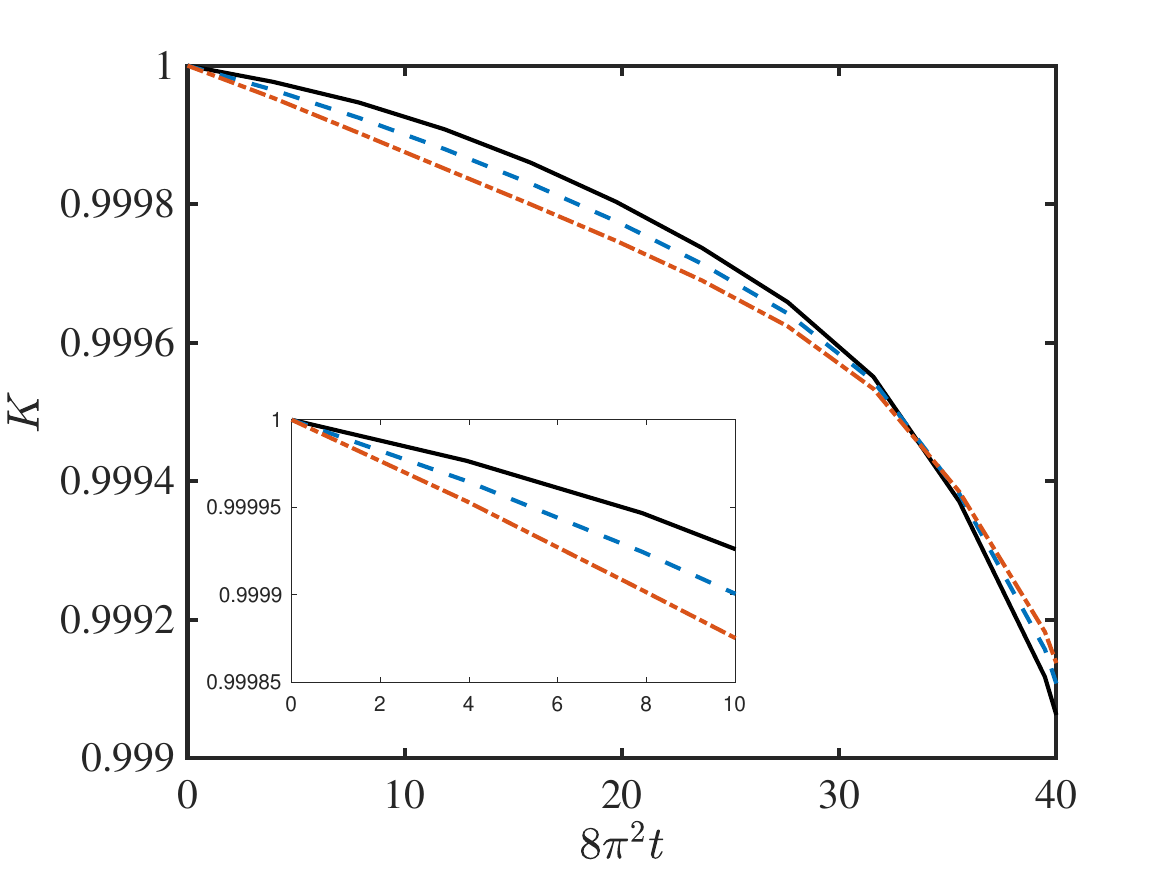}
    \caption{Global kinetic energy dissipation for three different constants $\alpha$ in Eq.~\eqref{eq:algorithm-momentum prediction}. Lines indicate the kinetic energy of $\alpha = 0$ (black solid line), $\alpha = 0.5$ (blue-dashed line), and $\alpha = 1.0$ (red-dash-dotted line). The density ratio is chosen as $\rho_{1}/\rho_{2} = 1000$. Local characteristic length $\Delta_{\text{local}}$ is used for the simulation.}
    \label{fig:app:pressure fraction}
\end{figure}

Global kinetic energy decay evolution for three different values of the coefficient $\alpha = 0.0$, $0.5$, and $1.0$, in Section~\ref{sec:projection method algorithm} are shown in Figure~\ref{fig:app:pressure fraction}.
The sensitivity of the decay rate to $\alpha$ is marginal. However, in the case with $\alpha = 0.0$ (which is used throughout the paper) results in the least dissipation of global kinetic energy during the early stages.

\section{Effect of density ratio on the global kinetic energy dissipation\label{sec:app:density ratio}}

\begin{figure}
    \centering
    \includegraphics[width=0.6\textwidth]{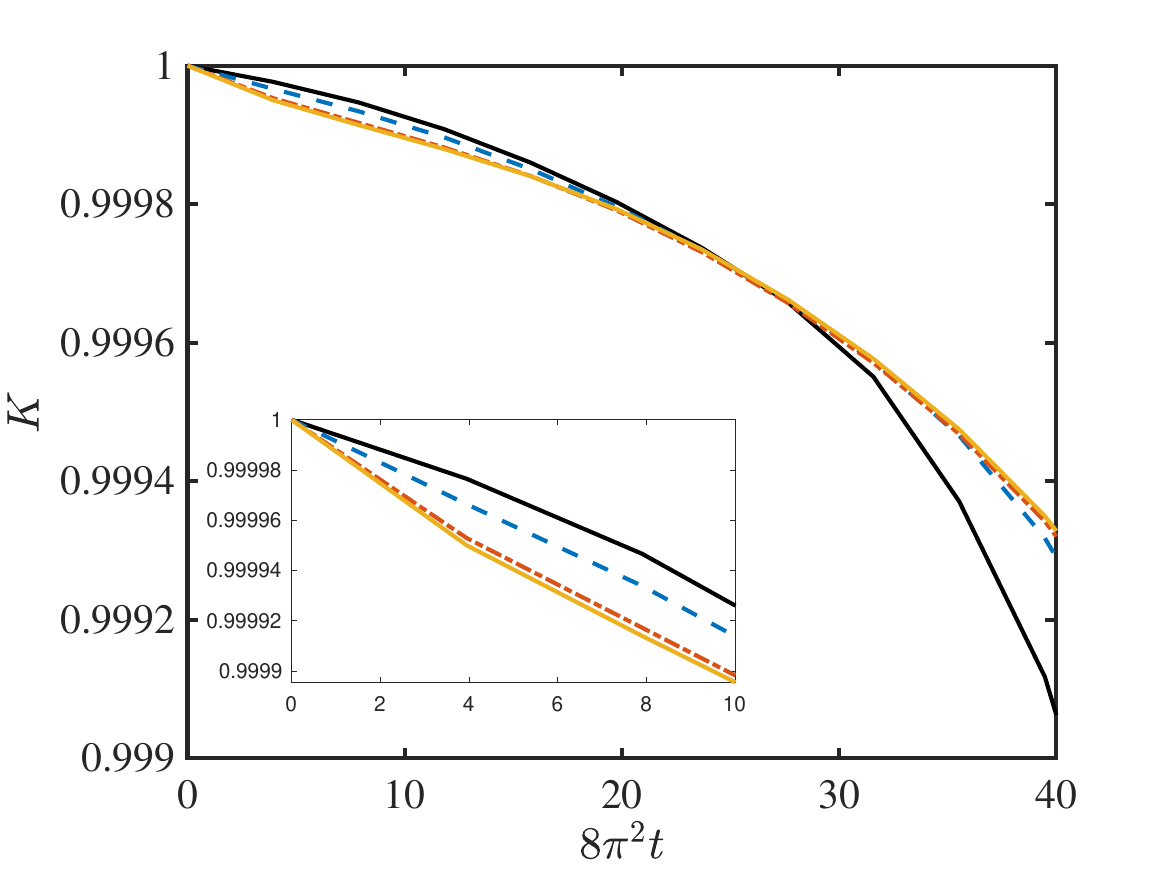}
    \caption{Global kinetic energy dissipation for different density ratios. Lines indicate the kinetic energy of $\rho_{1}/\rho_{2} = 1000$ (black solid line), $100$ (blue-dashed line), $10$ (red-dash-dotted line), and $1$ (yellow-solid line).}
    \label{fig:app:density ratios}
\end{figure}

The influence of the density ratio on the evolution of global kinetic energy is examined. Drop simulations in HIT field cases are conducted with four different density ratios: $\rho_1/\rho_2 = 1$, $10$, $100$, and $1000$. The results are presented in Figure~\ref{fig:app:density ratios}. Initially, the impact of the density ratio is minimal. The case with the highest density ratio, $\rho_1/\rho_2 = 1000$, exhibits the least dissipative behavior during the initial stage, as indicated by the factor $1/\rho_p$ in Eq.~\eqref{eq:energy budget: error analysis 4}. However, in the later stages of the simulation, the case with the highest density ratio shows the fastest decay of global kinetic energy, probably due to the larger pressure gradients induced due to higher inertia.

\bibliographystyle{model1-num-names}
\bibliography{two_phase}

\end{document}